\documentclass[twocolumn,trackchanges]{aastex631}

\begin{document}

\title{Magnetic fields in mini starburst complex Sgr B2}

\author[0000-0003-1337-9059]{Xing Pan}
\affiliation{School of Astronomy and Space Science, Nanjing University, 163 Xianlin Avenue, Nanjing 210023, P.R.China}
\affiliation{Key Laboratory of Modern Astronomy and Astrophysics (Nanjing University), Ministry of Education, Nanjing 210023, P.R.China}
\affiliation{Center for Astrophysics $\vert$ Harvard \& Smithsonian, 60 Garden Street, Cambridge, MA, 02138, USA}
\author[0000-0003-2384-6589]{Qizhou Zhang}
\affiliation{Center for Astrophysics $\vert$ Harvard \& Smithsonian, 60 Garden Street, Cambridge, MA, 02138, USA}
\author[0000-0002-5093-5088]{Keping Qiu}
\affiliation{School of Astronomy and Space Science, Nanjing University, 163 Xianlin Avenue, Nanjing 210023, P.R.China}
\affiliation{Key Laboratory of Modern Astronomy and Astrophysics (Nanjing University), Ministry of Education, Nanjing 210023, P.R.China}
\author[0000-0002-1407-7944]{Ramprasad Rao}
\affiliation{Center for Astrophysics $\vert$ Harvard \& Smithsonian, 60 Garden Street, Cambridge, MA, 02138, USA}
\author[0000-0001-6924-9072]{Lingzhen Zeng}
\affiliation{Center for Astrophysics $\vert$ Harvard \& Smithsonian, 60 Garden Street, Cambridge, MA, 02138, USA}
\author[0000-0003-2619-9305]{Xing Lu}
\affiliation{Shanghai Astronomical Observatory, Chinese Academy of Sciences, 80 Nandan Road, Shanghai 200030, P.R.China}
\author[0000-0002-4774-2998]{Junhao Liu}
\affiliation{National Astronomical Observatory of Japan, 2-21-1 Osawa, Mitaka, Tokyo, 181-8588, Japan}
%% Note that the \and command from previous versions of AASTeX is now
%% depreciated in this version as it is no longer necessary. AASTeX 
%% automatically takes care of all commas and "and"s between authors names.

%% AASTeX 6.31 has the new \collaboration and \nocollaboration commands to
%% provide the collaboration status of a group of authors. These commands 
%% can be used either before or after the list of corresponding authors. The
%% argument for \collaboration is the collaboration identifier. Authors are
%% encouraged to surround collaboration identifiers with ()s. The 
%% \nocollaboration command takes no argument and exists to indicate that
%% the nearby authors are not part of surrounding collaborations.

%% Mark off the abstract in the ``abstract'' environment. 
\begin{abstract}

We report the first arcsecond-resolution observations of the magnetic field in the mini starburst complex Sgr B2. SMA polarization observations revealed magnetic field morphology in three dense cores of Sgr B2 N(orth), M(ain), and S(outh). The total plane-of-sky magnetic field strengths in these cores are estimated to be 4.3-10.0 mG, 6.2-14.7 mG, and 1.9-4.5 mG derived from the angular dispersion function method after applying the correction factors of 0.21 and 0.5. Combining with analyses of the parsec-scale polarization data from SOFIA, we found that a magnetically supercritical condition is present from the cloud-scale ($\sim$10 pc) to core-scale ($\sim$0.2 pc) in Sgr B2, which is consistent with the burst of star formation activities in the region likely resulted from a multi-scale gravitational collapse from the cloud to dense cores.

\end{abstract}

%% Keywords should appear after the \end{abstract} command. 
%% The AAS Journals now uses Unified Astronomy Thesaurus concepts:
%% https://astrothesaurus.org
%% You will be asked to selected these concepts during the submission process
%% but this old "keyword" functionality is maintained in case authors want
%% to include these concepts in their preprints.

%% From the front matter, we move on to the body of the paper.
%% Sections are demarcated by \section and \subsection, respectively.
%% Observe the use of the LaTeX \label
%% command after the \subsection to give a symbolic KEY to the
%% subsection for cross-referencing in a \ref command.
%% You can use LaTeX's \ref and \label commands to keep track of
%% cross-references to sections, equations, tables, and figures.
%% That way, if you change the order of any elements, LaTeX will
%% automatically renumber them.
%%
%% We recommend that authors also use the natbib \citep
%% and \citet commands to identify citations.  The citations are
%% tied to the reference list via symbolic KEYs. The KEY corresponds
%% to the KEY in the \bibitem in the reference list below. 

\section{Introduction}
Star formation is predominantly influenced by the interplay between gravity, turbulence, and magnetic fields. Gravity drives the formation of high-density regions that may collapse to initiate star formation. Turbulence can either promote the formation of dense structures through shock compression or suppress the formation of dense gas by providing internal support \citep[e.g.,][]{2010A&A...512A..81F, 2020MNRAS.493.4643M}{}{}. In the past decade, significant progress has been made toward mapping the magnetic field topology in star forming regions thanks largely to advances in polarimetry equipped in far-infrared and radio telescopes. A variety of magnetic field morphology in high-mass star forming regions has been revealed by accumulated high-resolution polarization observations, ranging from an hourglass-like shape indicating strong magnetic fields regulating gas infall \citep{2014ApJ...794L..18Q,2019A&A...630A..54B,2021ApJ...923..204C} to a spiral-like pattern suggesting that gravity dominates the magnetic field \citep{2019A&A...630A..54B,2020ApJ...904..168B,2021ApJ...915L..10S,2022ApJ...941...51H}. While studies of individual regions suggest that magnetic fields may play a varying role in massive star formation\citep[][]{2019FrASS...6....3H,2023ASPC..534..193P}{}{}, surveys of a sample of massive star forming regions have found that the magnetic field is dynamically important during the fragmentation of parsec-scale massive molecular clumps \citep[][]{2014ApJ...792..116Z}{}{}.

In this paper, we explore the magnetic field and its dynamic role in a mini starburst region in Sagittarius B2 (hereafter Sgr B2). Such regions are characterized by intense star formation activities under conditions comparable to those observed in starburst galaxies \citep[][]{2003ApJ...582..277M, 2013ApJ...778...96W}{}{}. Despite some recent studies \citep[][]{2021ApJ...914...24L, 2023ApJ...953..113L} exploring the magnetic field in starburst galaxies, the role of magnetic fields on star formation in these more extreme environments remains unclear. Studying the galactic mini-starburst regions may provide insights into the role of magnetic fields in external galaxies. The object of interest, Sgr B2, is embedded in one of the most massive molecular clouds in the Galaxy, located at a distance of $8.34\pm0.16$ kpc \citep{2014ApJ...783..130R}. It also hosts one of the most active star-forming regions in the Galaxy, harboring one rare mini-starburst with numerous massive (proto)stars and HII regions \citep[][]{1995ApJ...449..663G,2011A&A...530L...9Q,2018ApJ...853..171G}{}{}. With its high star formation rate ($\sim0.01M_\odot\mathrm{yr^{-1}}$) and efficiency \citep[][]{2017A&A...603A..89K,2017MNRAS.469.2263B,2018ApJ...853..171G, 2019ApJS..244...35L}, Sgr B2 is in contrast with the quiescent Central Molecular Zone (CMZ) clouds \citep{2013MNRAS.429..987L}, in which a large amount of dense molecular gas is present. \cite{2018ApJ...853..171G} and \cite{2022ApJ...929...34M} investigated the dynamical state of Sgr B2 cores, and found that most of the cores are gravitationally bound, which is consistent with the active star formation in the Sgr B2 region. However, the analysis did not consider the magnetic field which provides additional internal support against gravitational collapse. Previous studies of magnetic fields in Sgr B2 only resolve large-scale ($>$1 pc) structures \citep[e.g.,][]{1995ApJ...449..663G,1997ApJ...487..237D,1997ApJ...487..320N,2010ApJS..186..406D,2023arXiv230601681B}{}{}. Here we present the first arcsecond-resolution ($\sim$0.1 pc) SMA polarization observations of three active star forming cores, Sgr B2 N(orth), Sgr B2 M(ain) and Sgr B2 S(outh) \citep[][]{1993A&A...280..208G}, to investigate the role of magnetic fields in Sgr B2. We analyze the SMA full polarization data in conjunction with the magnetic field data within the 10 pc region of Sgr B2 obtained with the SOFIA HAWC+ \citep{2023arXiv230601681B}. The paper is organized as follows: In Section \ref{sec:obs}, we summarize the SMA and SOFIA observations. In Section \ref{sec:results}, we present the magnetic field maps and derive the field strengths. Discussions of the role of magnetic fields and dynamic states of the cores in Sgr B2 are presented in Section \ref{sec:discussion}. In Section \ref{sec:summary}, we provide a summary of the paper. 

\begin{table*}[!htp]
    \centering
    \caption{List of Observational Parameters}
    \label{tab:obsinfo}
    \begin{tabular}{lccclllll}
    \hline
    \hline
    Source & $\alpha$ & $\delta$ & Observation & Array\tablenotemark{b} & Bandpass & Gain & Flux & Beam\tablenotemark{c} \\
    & (J2000) & (J2000) & date\tablenotemark{a} & Configuration & Calibrator & Calibrator & Calibrator &  \\
    \hline
    Sgr B2(N) & 17:47:19.88 & -28:22:18.40 & 220609 & Extended (6) & 3C279 & 1733-130 & Callisto& $1.68\arcsec\times1.26\arcsec$ \\
    & & & 220722 & Compact (6) & 3C84 & 1733-130 & Callisto & \\
    & & & 220802 & Compact (6) & 3C84 & 1733-130 & Titan &\\
    Sgr B2(M) & 17:47:20.16 & -28:23:05.00 & 220610 & Extended (6) & 3C279 & 1733-130 & Callisto& $1.49\arcsec\times1.14\arcsec$\\
    & & & 220725 & Compact (6) & bllac & 1733-130 & Callisto & \\
    Sgr B2(S) & 17:47:20.19 & -28:23:45.00 & 220821 & Compact (6) & bllac & 1733-130 & Callisto& $3.57\arcsec\times2.12\arcsec$\\
    \hline
    \end{tabular}
    \tablecomments{\tablenotemark{a} Data of Observations is listed in a format of \emph{yymmdd}. \tablenotemark{b} The number in the parentheses denotes the number of available antennas during each observation. \tablenotemark{c} Synthesized beam combining all configurations for each core.}
\end{table*}

\section{Observations and data reduction}\label{sec:obs}
\subsection{SMA observations}
SMA\footnote{The Submillimeter Array is a joint project between the Smithsonian Astrophysical Observatory and the Academia Sinica Institute of Astronomy and Astrophysics, and is funded by the Smithsonian Institution and the Academia Sinica.} polarization observations (Project ID: 2022A-S018, PI: Zhang) of Sgr B2 were carried out in six tracks toward three clumps. Sgr B2 N(orth), M(ain), S(outh) from 2022 may through 2022 August. During the observations, the weather conditions were excellent with the typical Zenith atmospheric opacity at 225 GHz lower than 0.06. Detailed information about the observations is listed in Table \ref{tab:obsinfo}. The receivers were tuned to cover the 334.3-346.3 GHz in the LSB and 354.3-366.3 GHz frequencies in the USB, respectively, with a uniform spectral resolution of 140 kHz ($\sim0.12~\mathrm{km~s^{-1}}$) per channel.

The basic data reduction procedures, including bandpass, gain, and flux calibrations, were done with the IDL MIR package\footnote{\url{https://lweb.cfa.harvard.edu/~cqi/mircook.html}}. The calibrated visibilities were then exported to MIRIAD for instrumental polarization calibration and imaging. 
% Our SMA data were corrected for instrumental polarization leakage with the MIRIAD task GPCAL. 
We estimate the instrumental polarization solutions to be accurate to 0.09\% (see details in Appendix \ref{appen:pol_leakage}). We jointly image the calibrated visibilities to produce Stokes \emph{I, Q, U} maps. The $1\sigma$ rms noise of the Stokes \emph{I} maps of the dust continuum emission reaches 0.3, 0.25, 0.09 Jy/beam for Sgr B2(N), (M), and (S), respectively, while the Stokes \emph{Q}/\emph{U} maps have rms noises of 5, 3 and 2.5 mJy/beam for Sgr B2(N), (M), and (S), respectively. The synthesized beam for each core is $1.68\arcsec\times1.26\arcsec$, $1.49\arcsec\times1.14\arcsec$ and $3.57\arcsec\times2.12\arcsec$, respectively. Because the polarized intensity and polarized percentage are deﬁned as positive values, the measured quantities of the two parameters are biased toward larger values \citep{2006PASP..118.1340V}. The debiased polarized emission intensity $P$ and fractional polarization $p$ are derived by:
\begin{equation}
    P=\sqrt{Q^2+U^2-\sigma^2_{Q/U}},
\end{equation}
and
\begin{equation}
    p=P/I,
\end{equation}
where $\sigma_{Q/U}$ is the 1$\sigma$ RMS noise of the Q/U maps. The polarization position angle is estimated by:
\begin{equation}
    \psi=\frac{1}{2}\tan^{-1}\left(\frac{U}{Q}\right),
\end{equation}
where Q and U are defined following the IAU convention where North corresponds to an angle of 0$^\circ$ with $\psi$ increasing in the counterclockwise direction. Appendix \ref{appen:pol_leakage} also shows the fractional polarization maps, and most of the linear polarization detections are above the calibration uncertainty.

\subsection{SOFIA observations}
We also present SOFIA (Stratospheric Observatory for Infrared Astronomy) 214 $\mu$m polarization data of Sgr B2 from a recent SOFIA/HAWC+ \citep[][]{2018JAI.....740011T,2018JAI.....740008H}{}{} legacy survey of dust polarization in the CMZ: the SOFIA/HAWC+ Far-Infrared Polarimetric Large Area CMZ Exploration (FIREPLACE) Survey. The survey covered $\sim0.5^\circ$ of the CMZ, including Sgr B2 and other molecular clouds, with an angular resolution of $19.6\arcsec$. We refer readers to \cite{2023arXiv230601681B} and \citet{2024arXiv240105317P} for an overview and observation details for the entire survey.

% The product was level IV reduction and downloaded from the SOFIA archive\footnote{\url{https://irsa.ipac.caltech.edu/applications/sofia}}.

\begin{figure*}[!ht]
\gridline{\fig{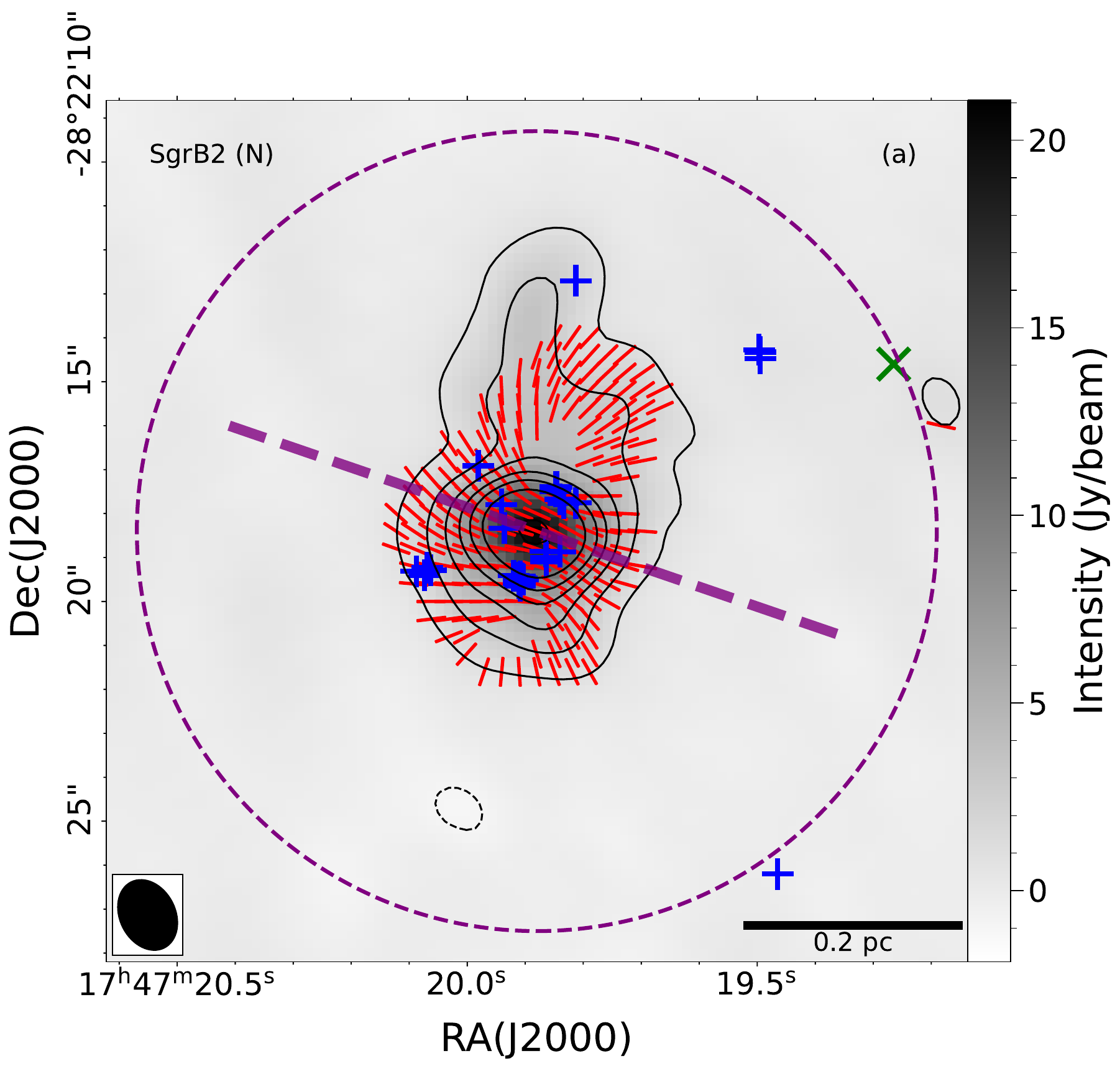}{0.33\textwidth}{}
          \fig{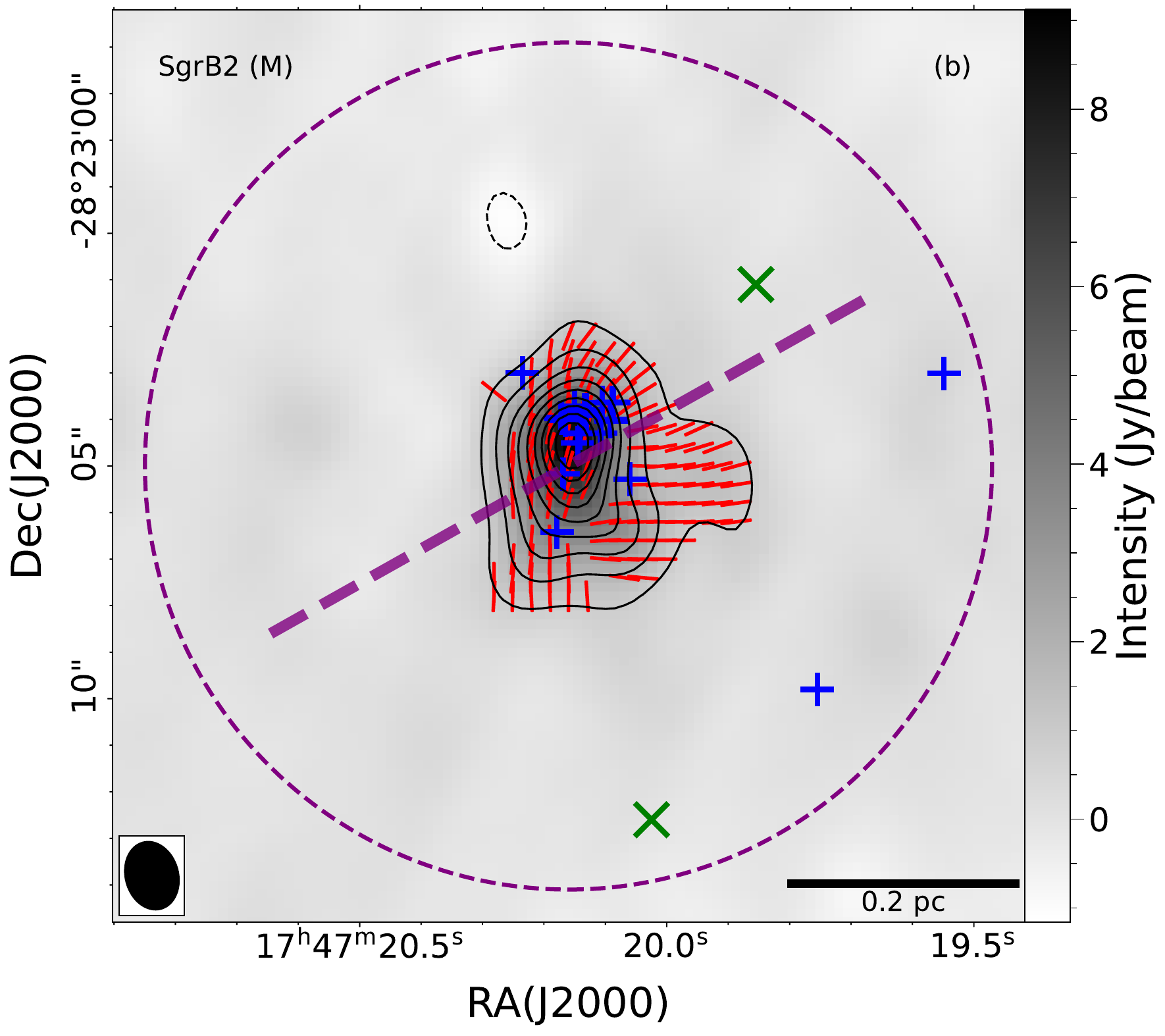}{0.325\textwidth}{}
          \fig{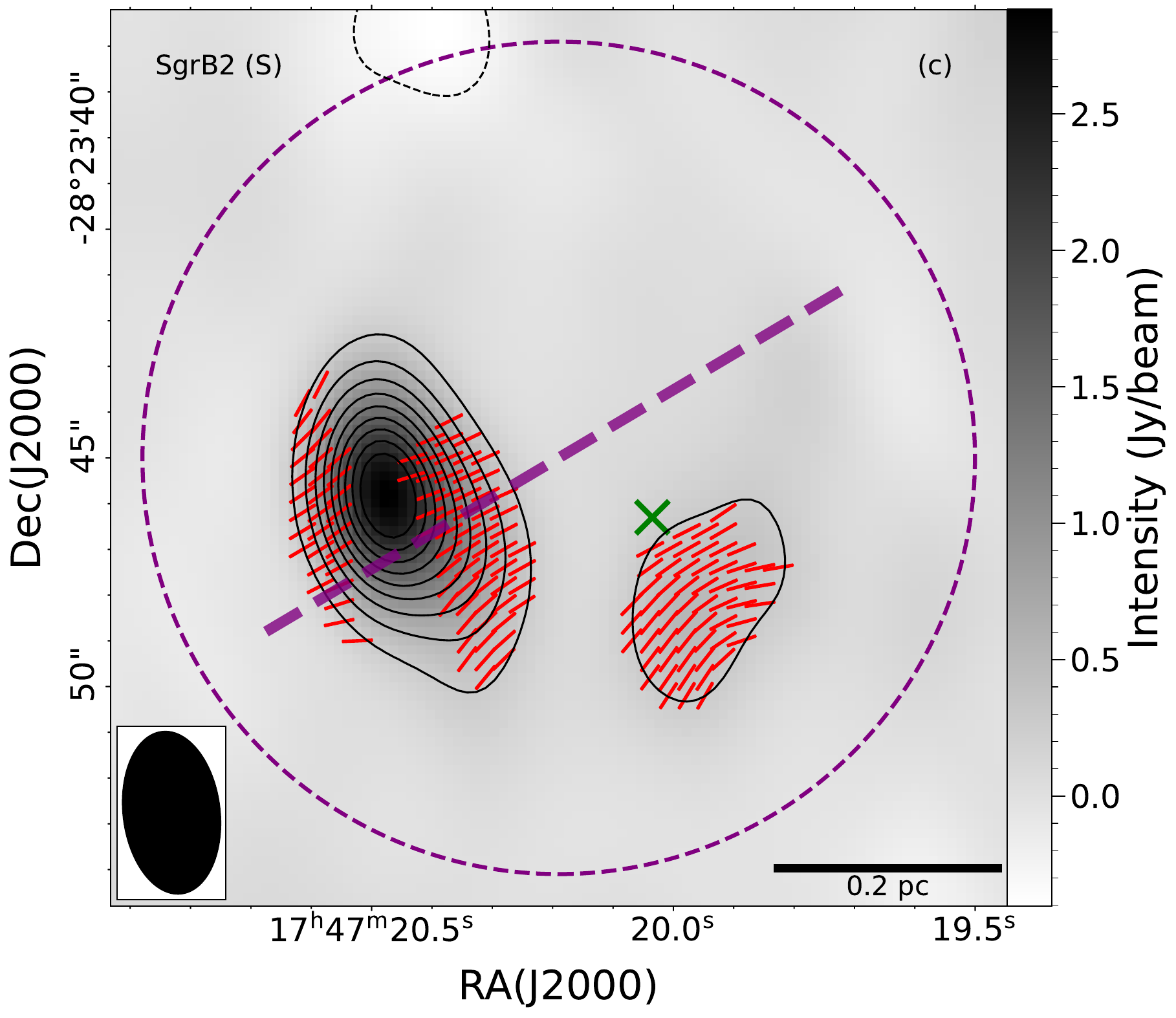}{0.33\textwidth}{}}
\gridline{
    \fig{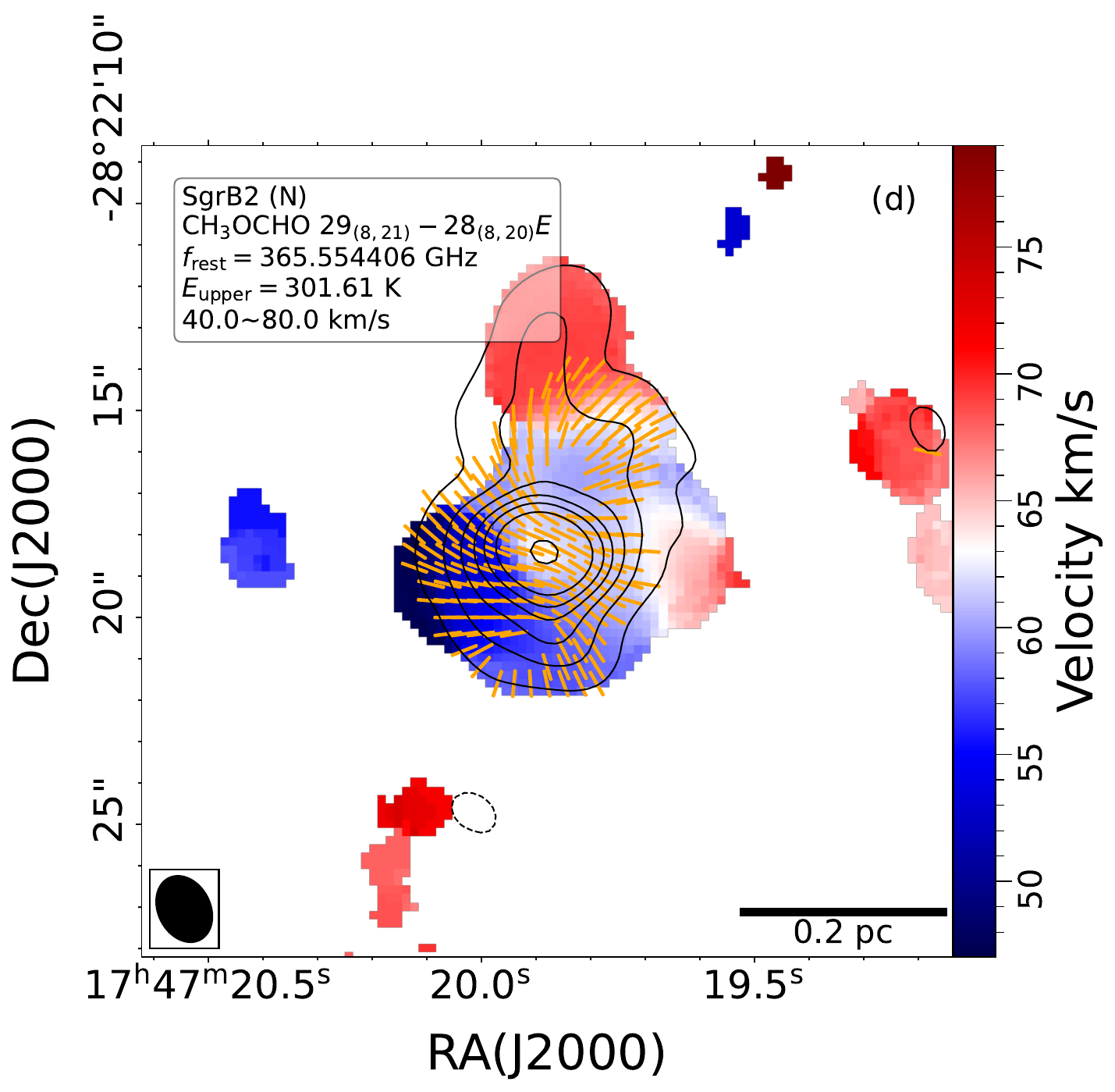}{0.33\textwidth}{}
    \fig{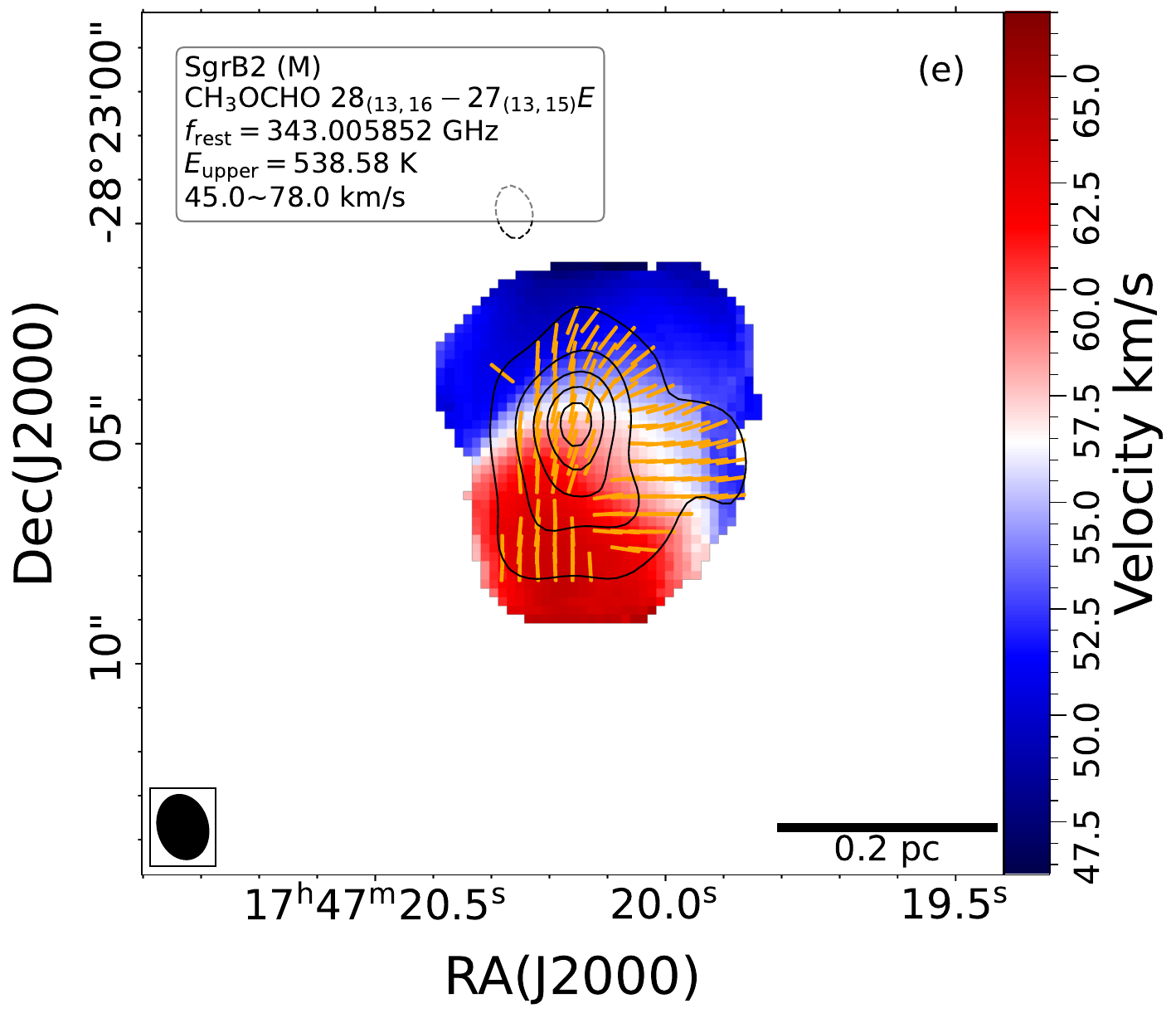}{0.33\textwidth}{}
    \fig{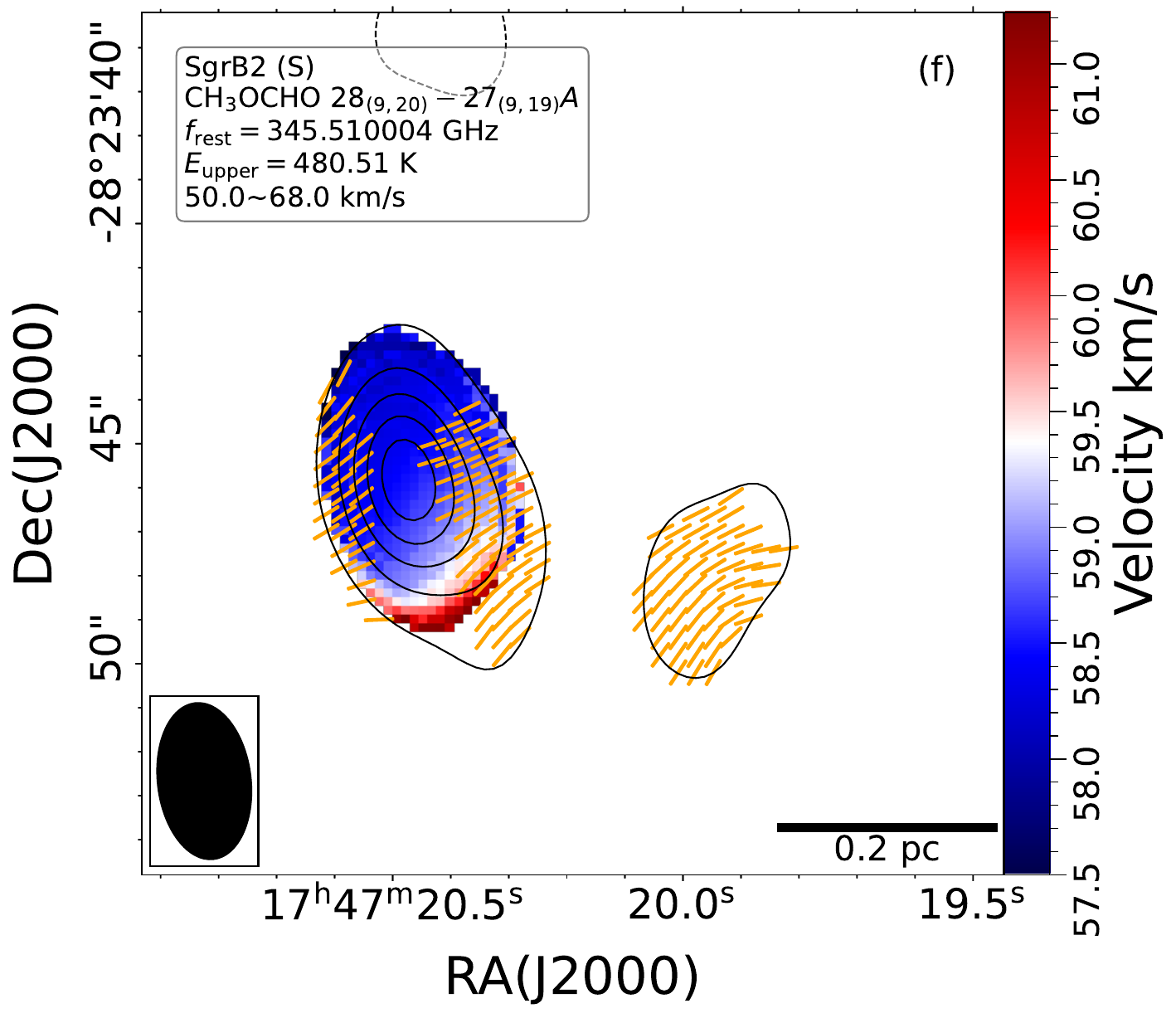}{0.33\textwidth}{}
    }
\gridline{
          \fig{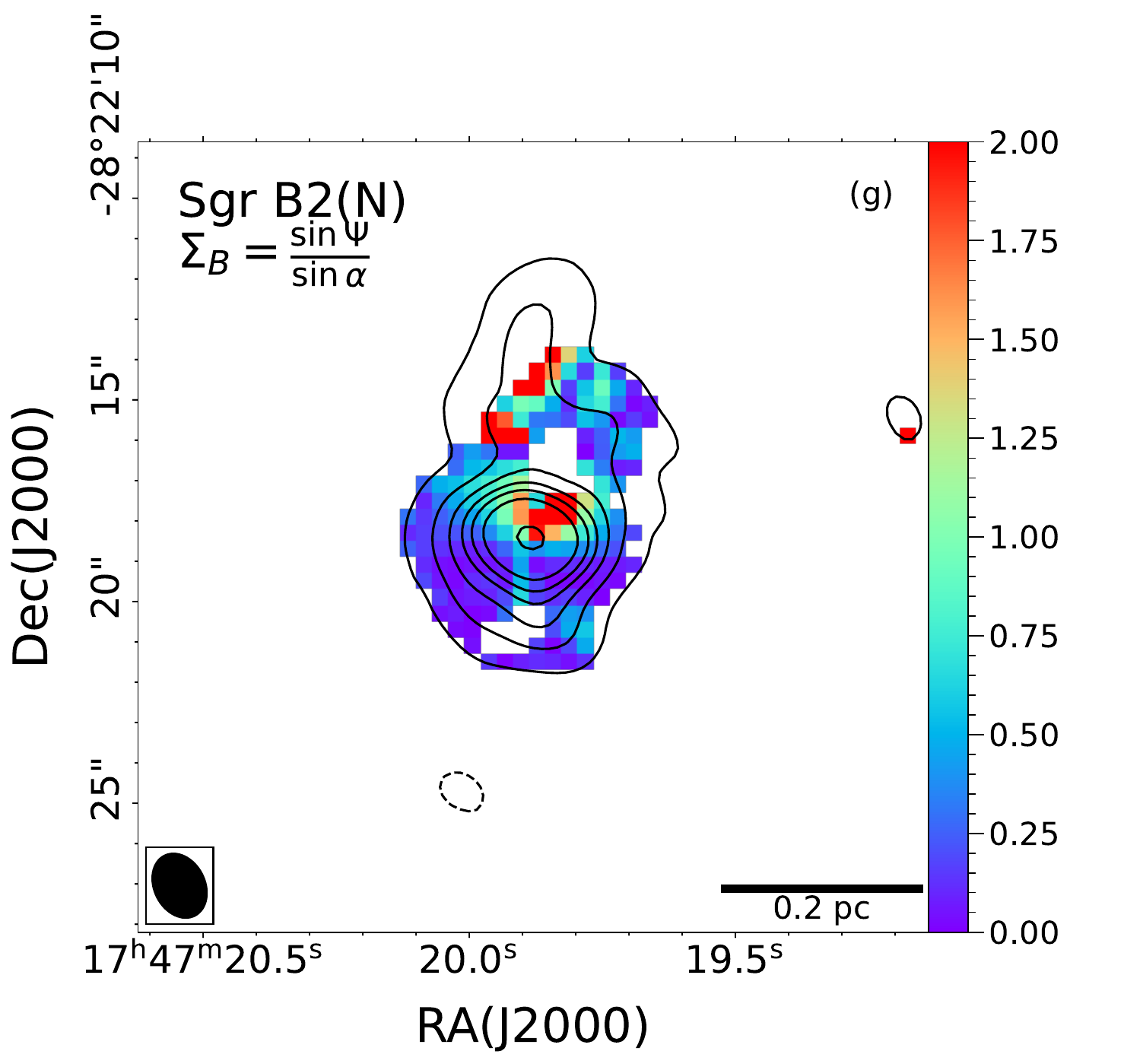}{0.33\textwidth}{}
          \fig{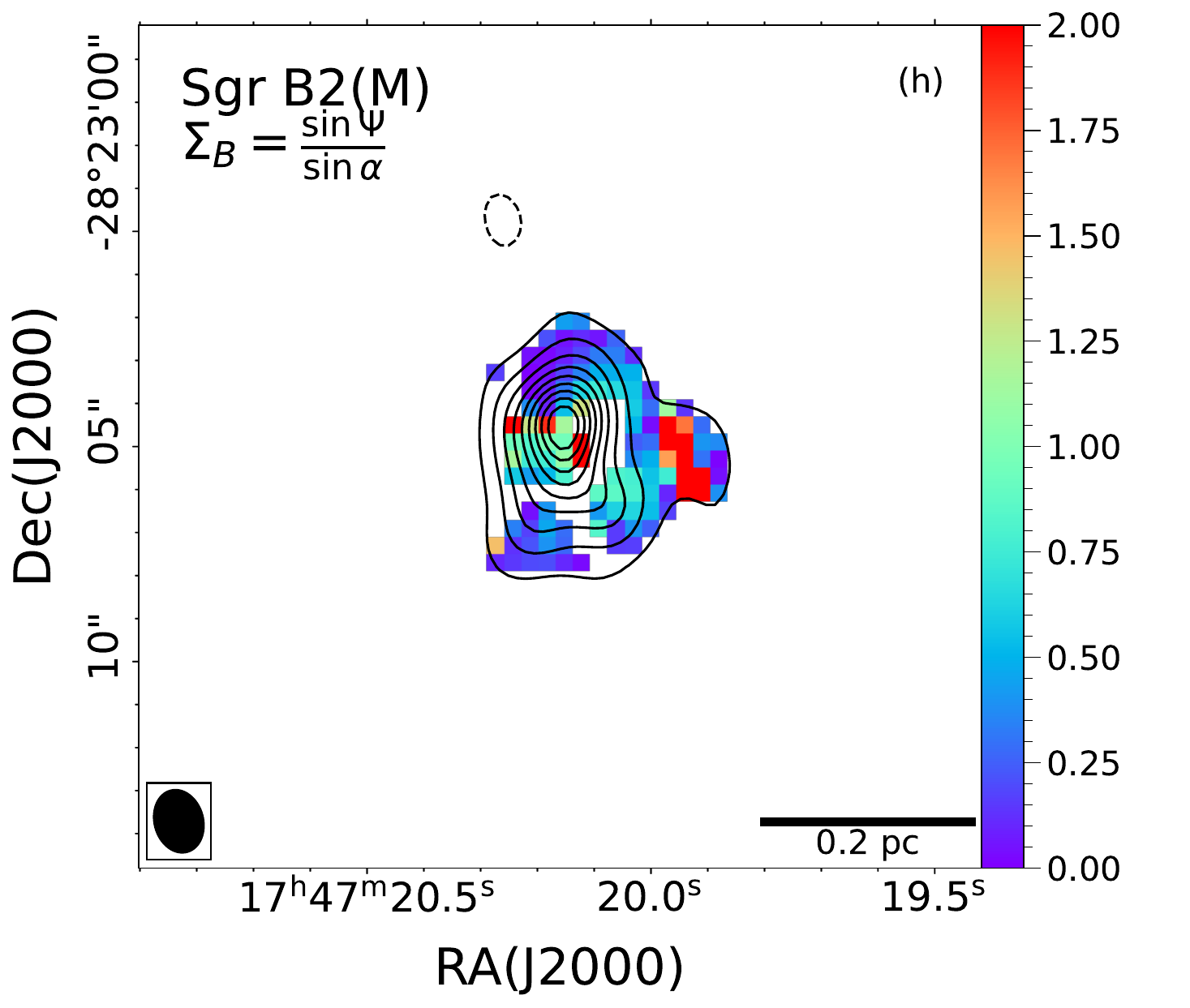}{0.33\textwidth}{}
          \fig{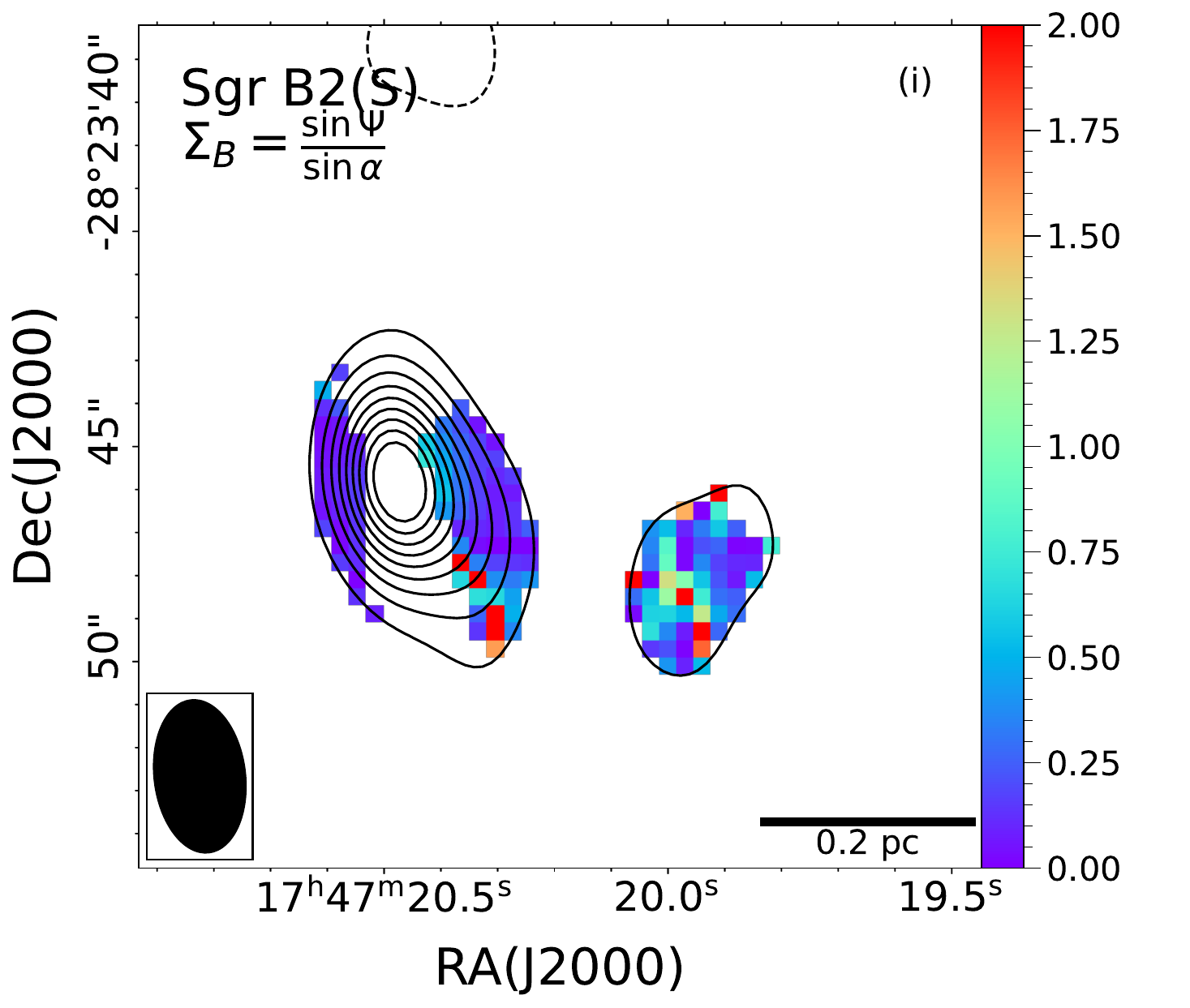}{0.33\textwidth}{}}
\caption{(a)-(c): The grayscale images and black contours show the Stokes \emph{I} of the SMA 1.3 mm continuum emission for Sgr B2 N(orth), M(ain), and S(outh). Contour levels are ($\pm 3, 6, 9, 12, 15, 18, 21, 24, 27, 30, 60, 90, 120, 150, 180)\times\sigma_I$, where $\sigma_I$ is the rms of Stokes \emph{I} data. Red segments show the orientation of the plane-of-sky magnetic field and have arbitrary length. Purple dashed circles and segments represent the $19.6\arcsec$ beam and the orientation of the magnetic ﬁeld from the SOFIA observations (see Fig. \ref{fig:sgrb2_sofia}) to compare with the SMA observations. $\mathrm{H_2O}$ \citep[][]{2004ApJS..155..577M,2014MNRAS.442.2240W}{}{} and $\mathrm{CH_3OH}$ masers \citep[][]{2010MNRAS.404.1029C}{}{} are marked by `$+$' and `$\times$' symbols, respectively. (d)-(f): Velocity field of the $\mathrm{CH_3OCHO}$ transitions overlaid with contours of Stokes \emph{I} and magnetic field morphology. The colored images show the intensity-weighted mean velocity (moment1) of the $\mathrm{CH_3OCHO}$ transitions. Orange segments indicate the orientation of the magnetic field, and black contours show the Stokes \emph{I} emission (same as (a)-(c)). The information of the $\mathrm{CH_3OCHO}$ transitions is labeled on the upper left corner of each panel. (g)-(i): Maps of $\Sigma_\mathrm{B}$. Contour levels are the same as in (a)-(c).}
\label{fig:SgrB2_mag}
\end{figure*}

\section{Results} \label{sec:results}
% \subsection{Magnetic Field Morphology}
\subsection{Continuum emission and magnetic fields from the SMA}
We observed the three sources (North, Main, and South) in Sgr B2 using the compact and extended configurations of the SMA. In Figs. \ref{fig:SgrB2_mag} (a)-(c), we present images of the 345 GHz dust continuum emission (Stokes \emph{I}) of these three objects ($\sim1$ pc). The continuum emission of these sources is resolved down to the core ($\sim0.1$ pc) scale. \citet{2016A&A...588A.143S} modeled the spectral energy distribution of Sgr B2(N) and (M), covering a wide range of frequencies from 40 GHz to 4 THz. They found that the contributions of the free-free emission in Sgr B2(N) and (M) are about 0.1 Jy/beam at 345 GHz with a beam width of 2$\arcsec$, while the peak intensities of these sources are about 10 Jy/beam. Therefore, the contributions of the free-free emission at 345 GHz are negligible. With the assumption of an optically thin dust emission, the gas mass of the dense core can be estimated with
\begin{equation}
\centering
M_{\mathrm{gas}}=\frac{gS_\nu D^2}{B_\nu(T_d)\kappa_\nu},
\end{equation}
% \begin{equation}
% \centering
% N_\mathrm{H_2}=\frac{M_\mathrm{gas}}{\mu_\mathrm{H_2}m_\mathrm{H}\pi \mathrm{FWHM^2_{mean}}},
% \end{equation}
% \begin{equation}
% \centering
% n_\mathrm{H_2}=\frac{3M_\mathrm{gas}}{4\mu_\mathrm{H_2}m_\mathrm{H}\pi \mathrm{FWHM^3_{mean}}},
% \end{equation}
where $S_\nu$ is the integrated flux above 3$\sigma$, D is the source distance which is set to 8.3 kpc, $B_\nu(T_d)$ is the Planck function at dust temperature $T_d$, $\kappa_\nu$ is the dust opacity. We assume a gas-to-dust mass ratio of 100 and the dust opacity $\kappa_\nu=10(\nu/1\mathrm{THz})^\beta$ in $\mathrm{cm^2\ g^{-1}}$ \citep{1983QJRAS..24..267H}. The dust emissivity index $\beta$ used here is 1.8 \citep{2016A&A...588A.143S}. Modified black body SED fittings from SCUBA \citep{2000ApJ...545L.121P} and \emph{Herschel} \citep{2013A&A...556A.137E} data both yield an average dust temperature of about 20 K for the moderate density region \citep[$\sim$5 pc scale, $n(H_2)\sim10^5\ \mathrm{cm^{-3}}$,][]{1991ApJ...369..157L,1993A&A...276..445H}{}{} in Sgr B2. However, higher temperatures are expected in the denser regions. \citet{2016A&A...588A.143S} modeled the dust temperature of Sgr B2 considering stars as heating sources. Their model shows that the central dust temperature of the two dense cores Sgr B2(N) and Sgr B2(M) can be over 300 K. Therefore, the two hot cores with higher densities should have dust temperatures higher than 20 K. \citet{2024ApJ...961....4B} argue that the two dense cores are more consistent with an average temperature of $\sim50$ K, based on their modeling. Here, we apply a dust temperature of 50 K to Sgr B2 (N) and (M). Since Sgr B2 (S) has a lower density and fewer star formation activities, it is not as strongly heated by embedded (proto)stars as the other two dense cores. The dust temperature profile of this core should be more flat. Thus, we take the average dust temperature for Sgr B2(S) of $\sim$30 K, estimated by a modified black body SED fitting at a larger scale \citep[$\sim1.5$ pc, from][]{2013A&A...556A.137E}. The average  $\mathrm{H_2}$ volume density of each core is measured above the $3\sigma$ contour of the continuum emission using an effective radius $R_\mathrm{eff}$:
\begin{equation}
    n_\mathrm{H_2}=\frac{3M_\mathrm{gas}}{4\mu_\mathrm{H_2}m_\mathrm{H}\pi R^3_\mathrm{eff}}
\end{equation}
where $R_\mathrm{eff}=\sqrt{A_\mathrm{target}/\pi}$, $\mu_{\mathrm{H_2}}=2.86$ is the mean hydrogen molecular weight \citep[][]{2013MNRAS.432.1424K, 2015MNRAS.450.1094P}{}{}. The derived gas mass and volume density of each dense core are listed in Table \ref{tab:adfpara}. For Sgr B2(S), our analysis focuses on the main component in the east.

In addition to the continuum emission, the SMA observations have revealed emission in many spectral lines from various molecules, including complex organic species \citep[][]{2008ApJ...677..353Q,2014ApJ...789....8N,2017A&A...604A..60B}{}{}. Analyzing such extensive spectral-line data is challenging, due to line blending effects caused by both known and unidentified molecular species, especially in a chemically rich region like Sgr B2. Moreover, the strong continuum emission in Sgr B2(N) and Sgr B2(M) may give rise to absorption features in certain transitions that further complicate the interpretation of line features. This paper focuses on the organic molecule $\mathrm{CH_3OCHO}$ (methyl formate). The frequency range of the SMA observations covers numerous $\mathrm{CH_3OCHO}$ lines, while we only select strong and non-blended lines for the analysis to avoid contamination from overlapping transitions from other molecules. Figs. \ref{fig:SgrB2_mag} (d)-(f) show the velocity maps of the three cores. Some show clear velocity gradients, implying internal bulk motions \citep[e.g.,][both detected a bipolar outflow nearly perpendicular to the north-south velocity gradient we detected in Sgr B2(N), suggesting rotation here]{2015ApJ...815..106H,2023arXiv231011339B}. We separate the small-scale turbulent fluctuations from the large-scale bulk motion. The derived nonthermal turbulent velocity dispersions are 4.4 km~s$^{-1}$, 6.4 km~s$^{-1}$, and 3.0 km~s$^{-1}$ for Sgr B2(N), (M), and (S), respectively. Further details on the determination of turbulent velocity dispersions can be found in Appendix \ref{appen:turbulence}.

\begin{figure*}[!ht]
    \gridline{
    \fig{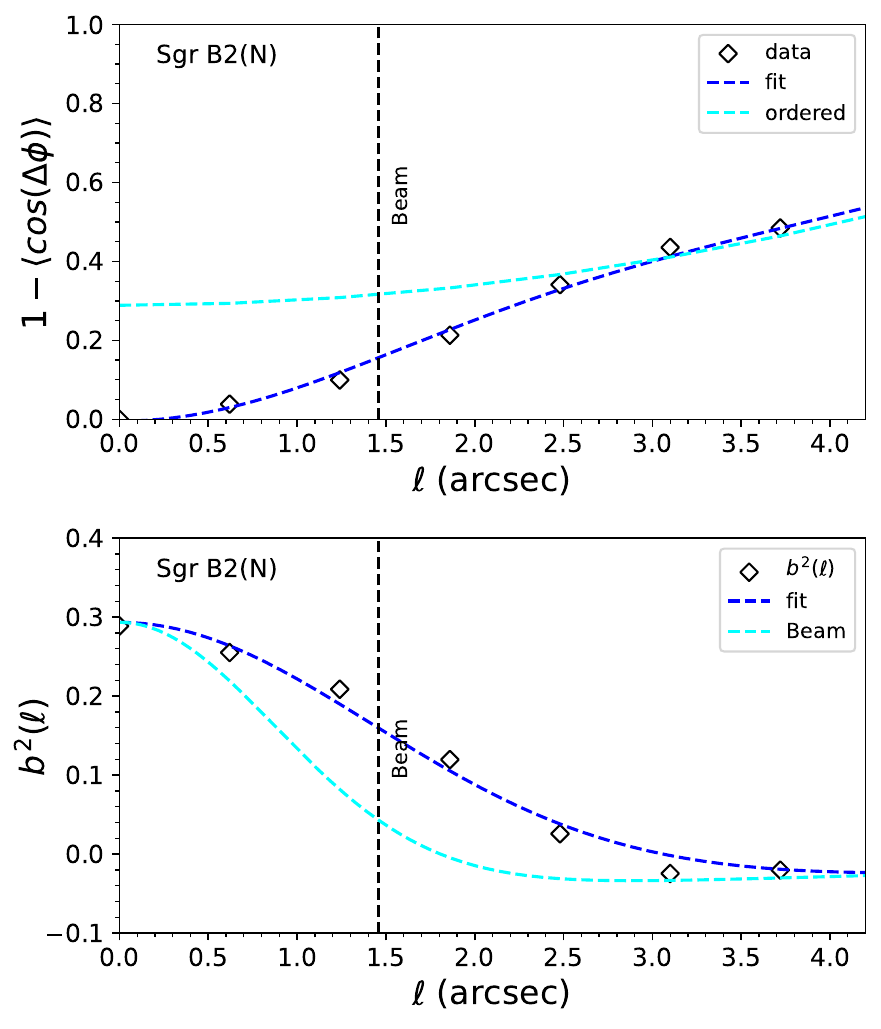}{0.45\textwidth}{(a)}
    \fig{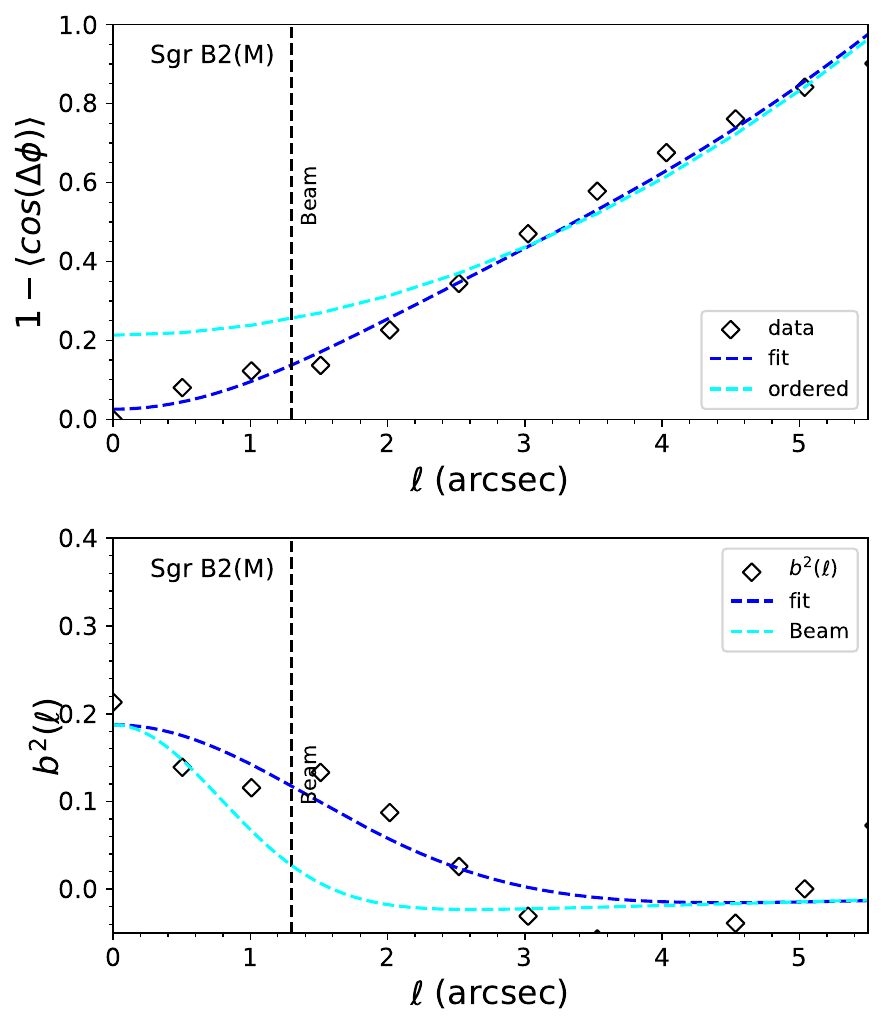}{0.45\textwidth}{(b)}
          }
    \gridline{
    \fig{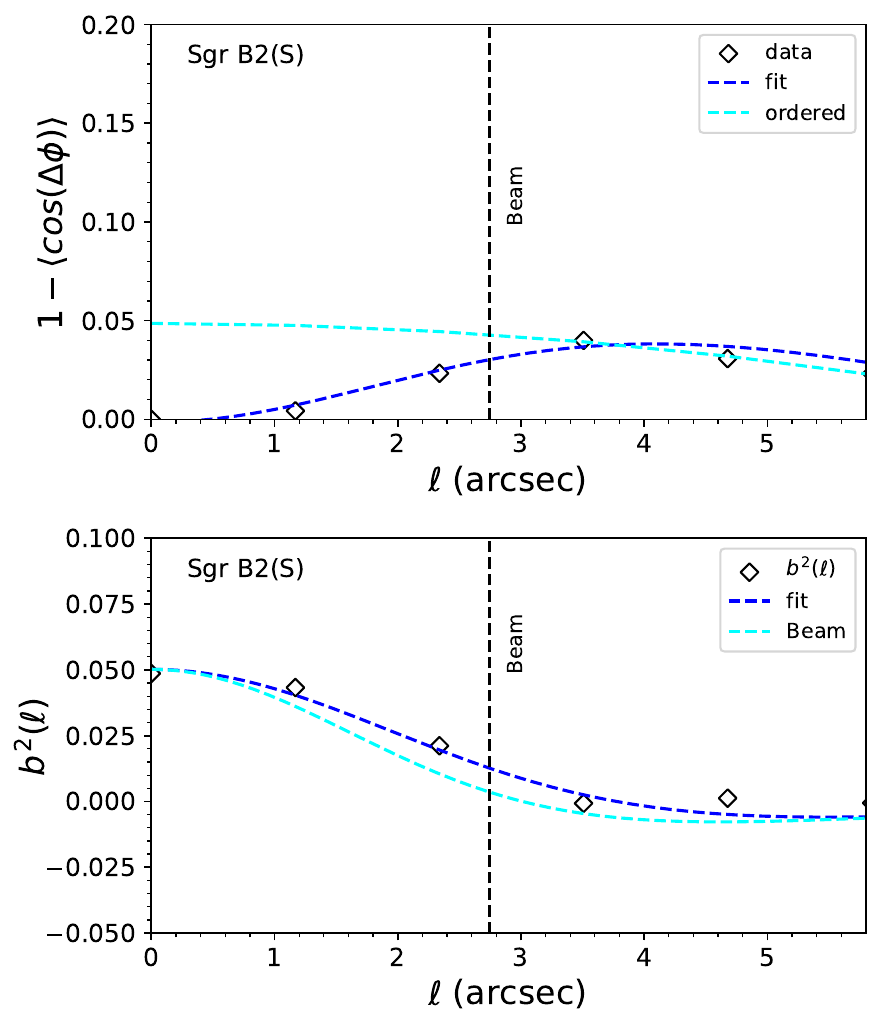}{0.45\textwidth}{(c)}
    \fig{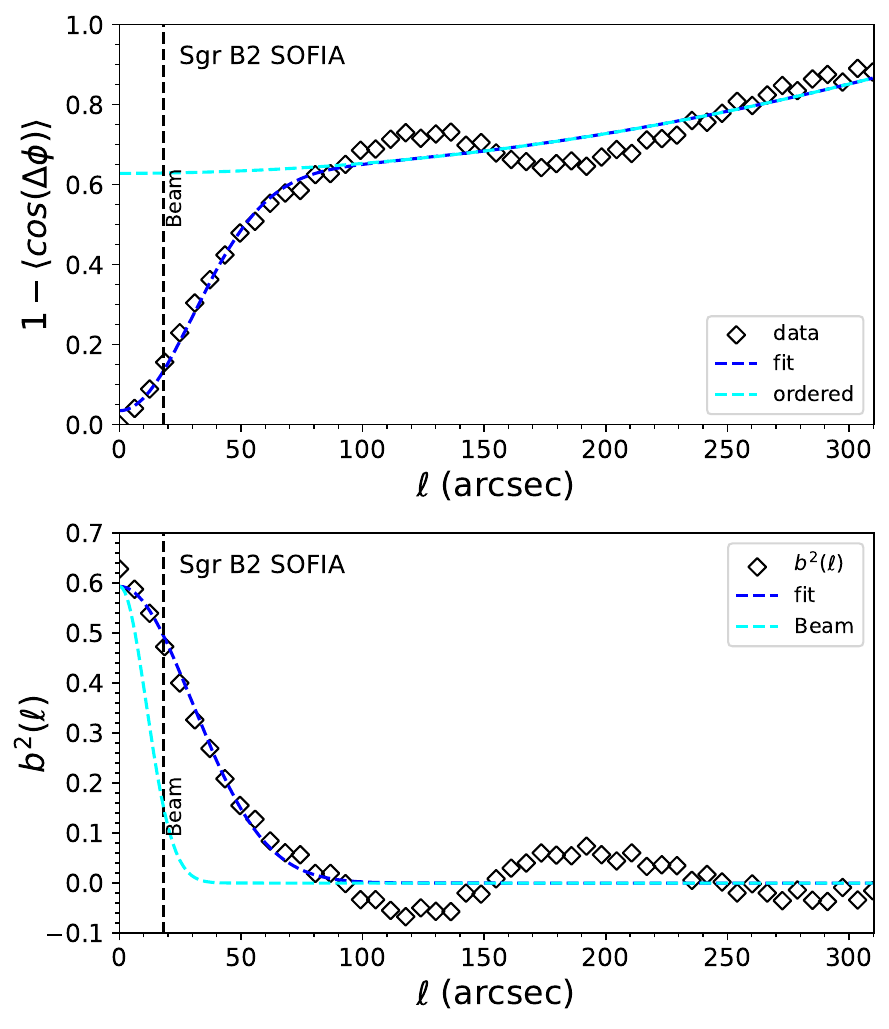}{0.45\textwidth}{(d)}
          }
    \caption{Angular dispersion function for the SMA data and SOFIA data.(a) Top: Diamond symbols show angle dispersion segments for Sgr B2(N). Blue dashed line shows the fitted angular dispersion function. Cyan dashed line shows the ordered component of the best fit. Vertical line indicates the beam size. Bottom: Turbulent component ($b^2(\ell)$) of the ADF method is shown in diamond symbols. Blue dashed line shows the best fit of $b^2(\ell)$. Cyan dashed line shows the correlated component distributed from the beam. (b)-(d): Same as (a) for Sgr B2(M), (S) and Sgr B2 from SOFIA data.}
    \label{fig:ADF_fit}
\end{figure*}

% \begin{figure*}[!ht]
%     \centering
%     \includegraphics[width=0.8\textwidth]{pol_fig/SgrB2_mag.pdf}
%     \caption{Maps of dust continuum and magnetic field morphology derived by dust polarization. (a): Red segments show the magnetic field of Sgr B2 obtained from SOFIA HAWC+ data. The background image and contours shows the SOFIA 214 $\mu$m continuum. Contours are drawn in steps of 15 $\sigma$, starting at 9$\sigma$ (for an rms value of 18 Jy $\mathrm{beam^{-1}}$). $\mathrm{H_2O}$ and $\mathrm{CH_3OH}$ masers are marked by `$+$' and `$\times$' symbols, respectively. The cyan circles mark the fields observed by SMA. (b)-(d): The gray scale images and black contours show the Stokes \emph{I} of SMA 1.3 mm continuum emission of Sgr B2 N(orth), M(ain), and S(outh). Contour levels are ($\pm 3, 6, 9, 12, 15, 18, 21, 24, 27, 30, 60, 90, 120, 150, 180)\times\sigma_I$, where $\sigma_I$ is the rms of Stokes \emph{I}. Red segments show the orientation of the plane-of-sky magnetic field and have arbitrary length. Purple dashed circles and segments represent the $18\arcsec$ beam and the orientation of the magnetic ﬁeld of the SOFIA observations to compare with the SMA observations. $\mathrm{H_2O}$ and $\mathrm{CH_3OH}$ masers are marked by `$+$' and `$\times$' symbols, respectively.}
%     \label{fig:SgrB2_mag}
% \end{figure*}

With the assumption that the shortest axis of dust grain is aligned with the magnetic field \citep[][]{2007MNRAS.378..910L,2007JQSRT.106..225L,2015ARA&A..53..501A}{}{}, we can obtain the plane-of-sky component of magnetic field orientations by rotating the orientation of the electric field by 90$^\circ$. We applied cuts in Stokes I intensity threshold of $I/\sigma_I>3$, and polarization intensity threshold of $P/\sigma_P>2$, to derive the magnetic field shown in Figs. \ref{fig:SgrB2_mag} (a)-(c). We also incorporated a polarization intensity threshold of $P/\sigma_P>3$, and tested it using the same magnetic field analysis method. We obtained results similar to those with a threshold of $P/\sigma_P>2$. To ensure sufficient detections for statistical analysis in each core, particularly in Sgr B2(S), we applied a threshold of $P/\sigma_P>2$ for SMA polarization observations. In Sgr B2(N), the magnetic field structure exhibits spiral-like features connected to the central core, roughly following orientations of the identified filaments F01, F04, F06, F08 in \citet{2019A&A...628A...6S}. On the other hand, the polarized emission in Sgr B2(M) is compact and centered at the continuum emission peak, showing a radially inward magnetic field structure. In Sgr B2(S), the magnetic field exhibits a well-ordered alignment along the southeast-northwest direction, appearing nearly uniform in both components. The core-scale magnetic field is aligned with the parental magnetic structure observed by SOFIA. Such an alignment is consistent with a dynamically strong magnetic field here.
\begin{table*}[!htp]
\centering
% \small
\caption{Physical Parameters Related in the ADF Analysis}
\begin{tabular}{lcccccccccc}
% \centering
% \small
\hline
\hline
Source & Mass & $R_\mathrm{eff}$ & $T_\mathrm{dust}$ & $n(H_2)$ & $\sigma_\mathrm{turb}$  & $\sqrt{\langle B^2_t \rangle / \langle B^2_0 \rangle}$ & $B^\mathrm{tot}_{pos}$\tablenotemark{d} & $\lambda$\tablenotemark{e} & $\mathcal{M}_A$\tablenotemark{f} & $\alpha_\mathrm{k+B}$\tablenotemark{g} \\
& $(10^3M_\odot)$ & (pc) & (K) & $(10^6\mathrm{cm}^{-3})$ & (km~s$^{-1}$)  & & (mG) & & & \\
\hline
Sgr B2(N) & 21.3 & 0.15 & 50\tablenotemark{a} & 20.0 & 4.4 & $1.2$ & $4.3$ & $17.7$ & 4.9 & $0.12\sim0.18$ \\
Sgr B2(M) & 10.2 & 0.12 & 50\tablenotemark{a} & 19.5 & 6.4  & $1.1$ & $6.2$ & $9.4$ & 3.3 & $0.36\sim0.59$ \\
Sgr B2(S) & 1.8 & 0.12 & 30\tablenotemark{b} & 3.4 & 3.0 & $0.6$ & $1.9$ & $5.4$ & 3.1 & $0.47\sim0.76$\\
\hline
Sgr B2 & 3250 & 6.1 & 20\tablenotemark{b} & 0.05 & 9.3\tablenotemark{c} & $2.8$ & $0.4$ & $20.3$ & 6.1 & $0.13\sim0.20$ \\
\hline
\end{tabular}
\label{tab:adfpara}
\tablecomments{\tablenotemark{a} \cite{2024ApJ...961....4B}. \tablenotemark{b} \cite{2013A&A...556A.137E}. \tablenotemark{c} \cite{2008MNRAS.386..117J}. \tablenotemark{d} The total plane-of-sky magnetic field strength applying correction factor of $Q_c=0.21$. \tablenotemark{e} Mass-to-flux ratio. \tablenotemark{f} Alfv$\mathrm{\acute{e}}$nic Mach number. \tablenotemark{g} Virial parameter with power-law index of density profile, $\rho\propto r^{-\beta}$, ranges from 0 to 2 ($\beta=0$ implies a uniform density profile and $\beta=2$ means a centrally peaked density profile.)} 
\end{table*}

The Davis-Chandrasekhar-Fermi method \citep[DCF method,][]{1951PhRv...81..890D, 1953ApJ...118..113C} and its variants are commonly used to estimate the strength of the plane-of-sky magnetic field ($B_\mathrm{pos}$), assuming that the perturbation in the magnetic field is driven by the turbulent motion. We refer readers to \citet{2022FrASS...9.3556L} for a detailed review of the DCF method. Further studies have expanded the DCF method using the angular dispersion function \citep[ADF,][]{2008ApJ...679..537F, 2009ApJ...706.1504H, 2009ApJ...696..567H,2016ApJ...820...38H} analysis to more accurately quantify the variations in the direction of the magnetic field, especially with an underlying ordered field structure. For the three sources in Sgr B2, the plane-of-sky magnetic field strength is given by \citep{2009ApJ...696..567H}:
\begin{equation}
    B_0=\sqrt{\mu_0\rho}\ \sigma_\mathrm{turb}\left[\frac{\langle B^2_t \rangle}{\langle B^2_0 \rangle} \right]^{-1/2}
\end{equation}
\begin{equation}
B^{tot}_{pos}= Q_c B_0\sqrt{1+\left[\frac{\langle B^2_t \rangle}{\langle B^2_0 \rangle} \right]}
\end{equation}
where $\mu_0$ is the vacuum permeability, $\rho=\mu_\mathrm{H_2}m_\mathrm{H_2}n_\mathrm{H_2}$ is the average mass density of gas, $\sigma_\mathrm{turb}$ is the turbulent velocity dispersion of the core, $\langle B^2_t \rangle/\langle B^2_0 \rangle$ is the turbulent-to-ordered magnetic energy ratio derived from the fitting result of the ADF analysis, $B^{tot}_{pos}$ is the total plane-of-sky magnetic field strength, and $Q_c$ is the correction factor. The details of ADF analysis are shown in Appendix \ref{appen:adfmethod}. To correct for the effect of smoothing in the magnetic fields, one needs the effective thickness ($\Delta^\prime$) of the observed region and the two Gaussian profiles of the synthesized beam ($W_1=\sqrt{\mathrm{FWHM_{maj}\times FWHM_{min}}}/\sqrt{8\ln2}$) and the large-scale filtering effect ($W_2$, derived by the shortest baseline of the array). Due to a limited number of detected polarization segments in the SMA data, we derived $\Delta^\prime$ of the three SMA cores by the ratio of its volume to the effective area of the core, $\Delta^\prime=V/A=(4/3)\pi r^3/(\pi r^2)=4/3r$ \citep{2023ApJ...954...99Z}, where r is the effective radius of the core. The $\Delta^\prime$ of Sgr B2(N), (M) and (S) are estimated to be $5.1\arcsec$, $4.0\arcsec$ and $3.9\arcsec$, respectively. Fig. \ref{fig:ADF_fit} shows the derived polarization angular dispersion data and the fitting results of the three cores. Because of the uncertainties in the assumption of the DCF method and the uncertainties in the statistics of magnetic field orientations, previous studies \citep[][]{2001ApJ...546..980O,2001ApJ...559.1005P,2021ApJ...919...79L}{}{} investigated the correction factor to correct for the magnetic field strength. Here, we applied the correction factor, $\bar{Q}_c=0.21$ from \citep{2021ApJ...919...79L}{}{}. They estimated the correction factor, $Q_c$, for the total plane-of-sky magnetic field strength derived by the ADF method in trans-Alfv$\mathrm{\acute{e}}$nic cores/clumps at 0.2$\sim$1 pc scales. With the parameters and correction factors outlined above, we obtained $B^\mathrm{tot}_\mathrm{pos}$ of $4.3$ mG, $6.2$ mG, and $1.9$ mG for Sgr B2(N), (M), and (S), respectively. Alternative correction factors for the DCF method were proposed. For instance, \citep{2001ApJ...546..980O} conducted MHD numerical simulations of molecular clouds and suggested $Q_c=0.5$. Similarly, \citep{2001ApJ...559.1005P} simulated three super-Alfv$\mathrm{\acute{e}}$nic cores, yielding an average correction factor of $Q_c=0.4$. When applying these correction factors, we obtained a range of $B^\mathrm{tot}_\mathrm{pos}$ values of 4.3-10.0, 6.2-14.7, and 1.9-4.5 mG for Sgr B2(N), (M), and (S), respectively.

We also applied a threshold of $P/\sigma_P>3$ to the data to derive the magnetic field strength for each core and obtained $B^\mathrm{tot}_\mathrm{pos}$ of 6.6, 8.2, and 1.6 mG, for Sgr B2(N), (M) and (S), respectively, using a correction factor of $Q_c=0.21$. These values are similar to those obtained with $P/\sigma_P>2$. Therefore, in order to include more polarization detections, in particular for Sgr B2(S) to ensure adequate statistics when fitting with the ADF method, we applied a selection criteria $P/\sigma_P>2$ for SMA polarization observations.

\begin{figure*}[!ht]
    \gridline{
    \fig{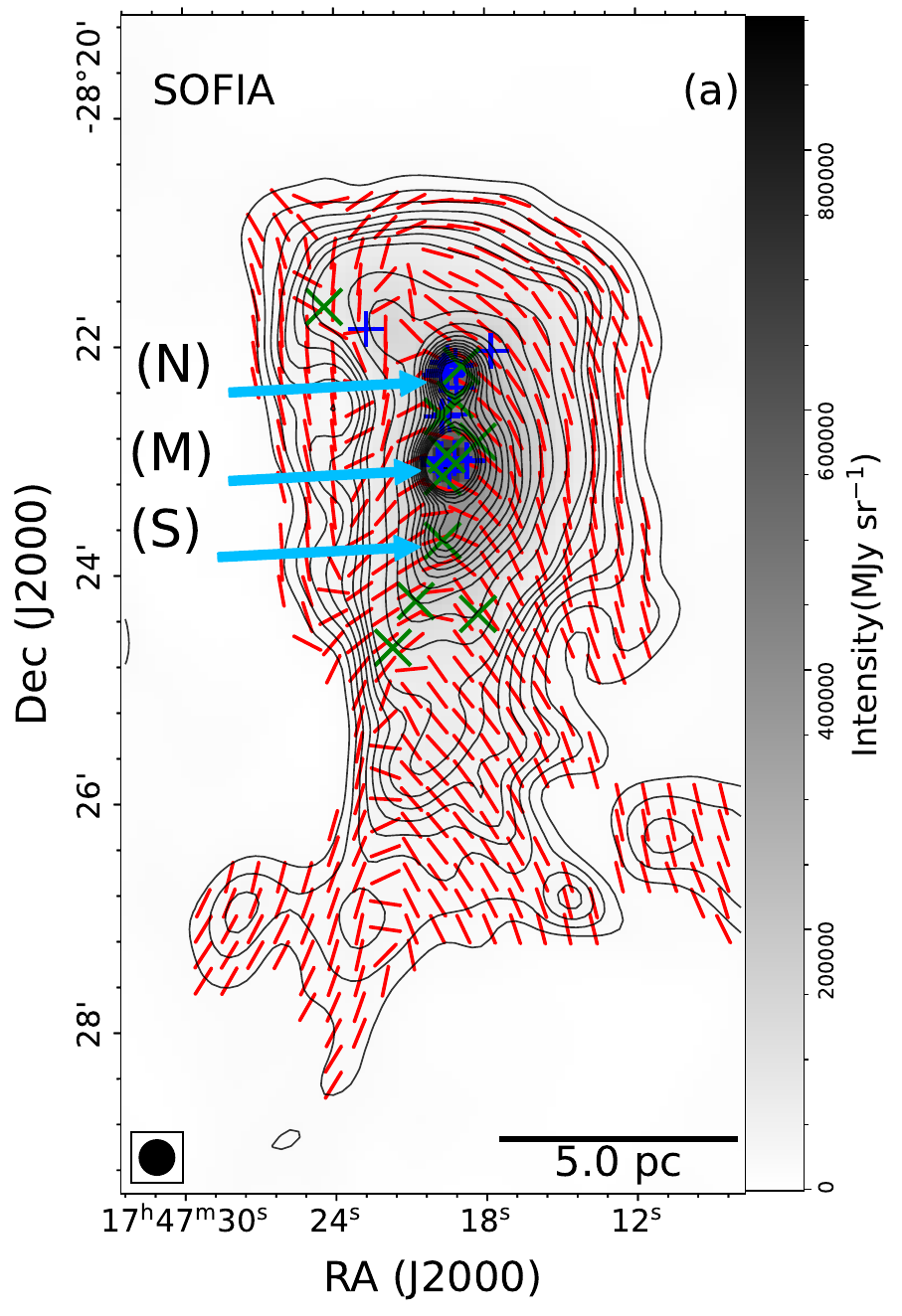}{0.45\textwidth}{}
    \fig{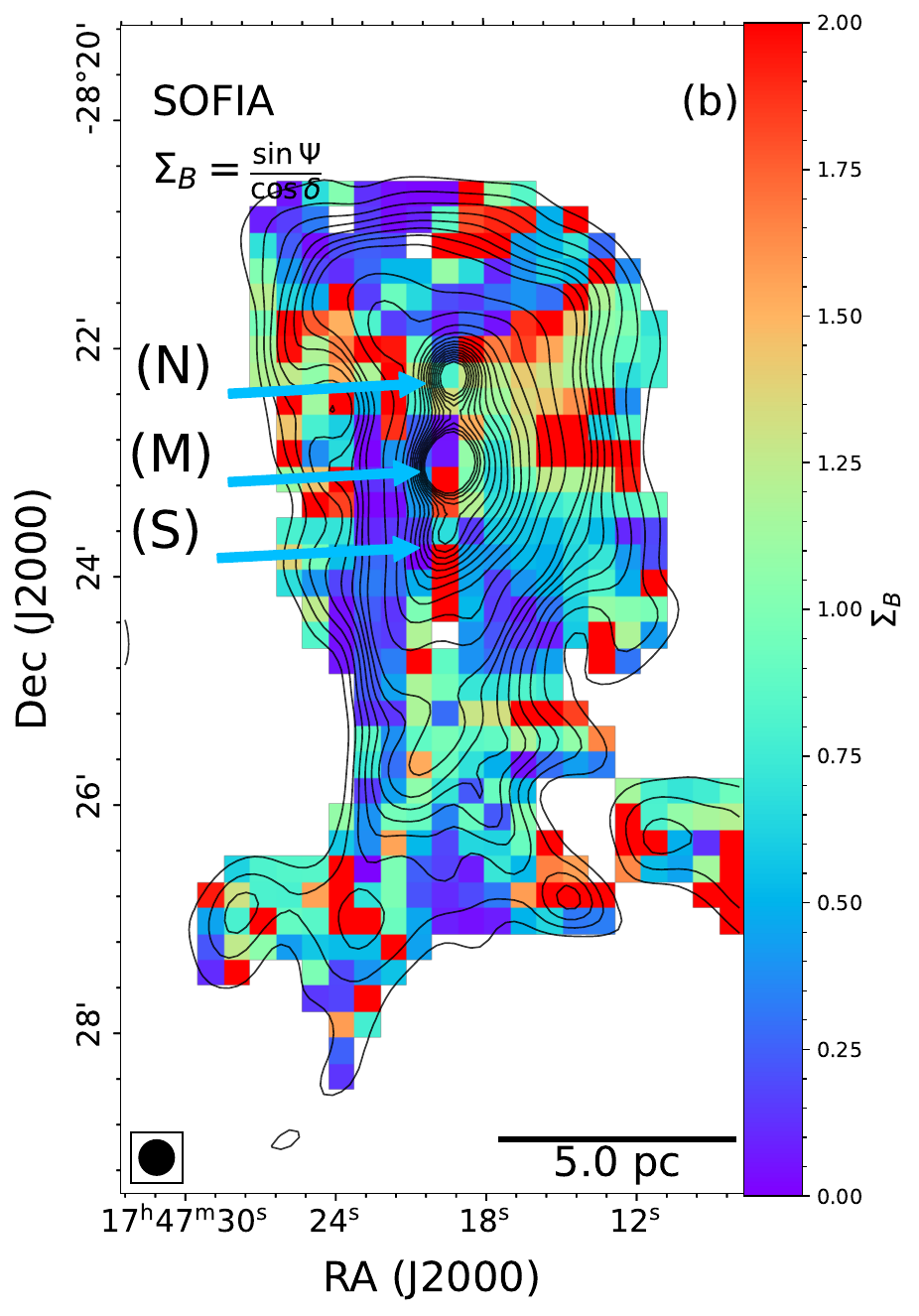}{0.46\textwidth}{}
    }
    \caption{(a): Magnetic fields (red line segments) of Sgr B2 obtained from the SOFIA HAWC+ data. The black contours show the SOFIA 214 $\mu$m Stokes I continuum emission. Contour levels are (800, 1100, 1400, 1700, 2000, 2300, 2600, 3000, 4500, 6000, 7500, 9000, 10500, 12000)$\times\sigma_I$, where $\sigma_I$ is the rms value of the SOFIA Stokes I intensity ($\sim$20 $\mathrm{MJy\ sr^{-1}}$). $\mathrm{H_2O}$ \citep[][]{2004ApJS..155..577M,2014MNRAS.442.2240W}{}{} and $\mathrm{CH_3OH}$ masers \citep[][]{2010MNRAS.404.1029C}{}{} are marked by `$+$' and `$\times$' symbols, respectively. (b): Map of $\Sigma_B$. Contour levels are the same as in (a).}
    \label{fig:sgrb2_sofia}

\end{figure*}

Assuming the intensity gradients can trace the direction of gas motions in the MHD force equation, \citet[][]{2012ApJ...747...79K} proposed a method (hereafter KTH method) to connect the relative directions among the magnetic field, intensity gradients, and local gravity to the magnetic field strength. The direction of local gravity can be derived assuming the dust emission is proportional to the gravitating mass and dust temperature is roughly constant over the three cores: 
\begin{equation}
    \mathbf{g_j(r)}=G\sum^{n}_{i=1}\frac{I_iI_j}{\bf{|r_i-r_j|^2}}\mathbf{e_{ij}}
\end{equation}
where $\mathbf{e_{ij}}$ is the unit direction vector between positions $\mathbf{r_i}$ and $\mathbf{r_j}$, $I_i$ and $I_j$ are the Stokes \emph{I} intensity at positions $\mathbf{r_i}$ and $\mathbf{r_j}$, respectively. We then derive the angular differences between the intensity gradient and local gravity orientation ($\psi$), the angular differences between the intensity gradient and the magnetic field ($\delta$), and the angular differences between the magnetic field and the local gravity orientation ($\omega$). With the assumption that the hydrostatic gas pressure ($F_P$) is negligible, the relative importance of the magnetic field force compared with gravity can be characterized by the ratio between local magnetic ﬁeld ($F_B$) to local gravity force ($F_G$) derived from $\psi$ and $\delta$:
\begin{equation}
    \Sigma_\mathrm{B}=\frac{F_B}{|F_G|}=\frac{\sin\psi}{\cos\delta}
\end{equation}
Figs. \ref{fig:SgrB2_mag} (g)-(i) show the resulting $\Sigma_\mathrm{B}$ maps for the three cores. The median values of $\Sigma_\mathrm{B}$ in Sgr B2(N), (M), and (S) are 0.45, 0.43, and 0.17, respectively. The small values of $\Sigma_\mathrm{B}$ indicate that the magnetic field in all these three cores is not strong enough to resist gravitational collapse.

\subsection{Continuum emission and magnetic field from SOFIA}\label{subsec:sofia_data}
The three massive dense cores detected by our SMA observations are embedded in a more extended, moderate density region \citep[$n(H_2)$ of $\sim10^5\ \mathrm{cm^{-3}}$ and $N(H_2)$ of $\sim10^{24}\ \mathrm{cm^{-3}}$,][]{1991ApJ...369..157L,1993A&A...276..445H}{}{}. Fig. \ref{fig:sgrb2_sofia} (a) shows the 214 $\mu$m continuum emission and the magnetic field of the moderate density region in Sgr B2 from SOFIA HAWC+ data, which covers a $10\times5$ pc area around the Sgr B2 complex. We applied cuts in the Stokes I intensity threshold of $I/\sigma_I>800$, percentage polarization less than 50\%, and polarization intensity threshold of $P/\sigma_P>3$, to derive the magnetic field shown in Fig. \ref{fig:sgrb2_sofia} (a). The magnetic field from the SOFIA observations is more uniform throughout the three dense clumps while exhibiting considerable variations outside these clumps. The average volume density $n_\mathrm{H_2}$ is measured over $800\sigma$ contour of the SOFIA Stoke I  emission. The estimated $n_\mathrm{H_2}$ is about $4.8\times10^4\ \mathrm{cm^{-3}}$. The average dust temperature is around 20 K from \citet{2013A&A...556A.137E} and \citet{2016A&A...588A.143S}. \citet{2008MNRAS.386..117J} undertook a 3-mm spectral-line imaging survey toward Sgr B2 using Mopra (a 22 m radio telescope of the Australia Telescope National Facility) data. They fit different lines detected in Sgr B2 and obtained a mean velocity width of about 22 $\mathrm{km~s^{-1}}$. Thus, the average velocity dispersion of Sgr B2 is 9.3 $\mathrm{km~s^{-1}}$. However, we take it as an upper limit of turbulence because of the coarser resolution ($\sim 35\arcsec$) of Mopra, which may not resolve underlying systemic motions. We applied the ADFs analysis (see Appendix \ref{appen:adfmethod}) to the SOFIA data over $\ell<6\arcmin$. The effective thickness ($\Delta^\prime$) of the cloud probed by the SOFIA observations is determined as the width at the half maximum in a normalized auto-correlation function of the integrated normalized polarization flux \citep{2009ApJ...706.1504H}. We obtained $\Delta^\prime$ of the SOFIA data of about $5.2\arcmin$. The derived polarization angular dispersion data and the fitting results of the ADF analysis are also shown in Fig. \ref{fig:ADF_fit}. The total plane-of-the-sky magnetic field strength for Sgr B2 derived from the SOFIA data is about $0.4$ mG using a correction factor $Q_c=0.21$, and the magnetic field strength ranges from 0.4 (assuming $Q_c=0.21$) to 0.9 mG (assuming $Q_c=0.5$).

The KTH method is also applied to constrain the significance of the large-scale magnetic field over gravity. Fig. \ref{fig:sgrb2_sofia} (b) shows the $\Sigma_B$ map derived from the SOFIA data. Most $\Sigma_B$ values are lower than 1. The median value of $\Sigma_B$ is 0.74, which implies that the large-scale magnetic field in Sgr B2 is also dominated by gravity.

\section{Discussion} \label{sec:discussion}
\subsection{Starburst in Sgr B2}
As we mentioned above, Sgr B2 undergoes a mini-starburst with a high star formation rate \citep[$\sim10^{-2}\ \mathrm{M_\odot yr^{-1}}$,][]{2017A&A...603A..89K,2017MNRAS.469.2263B,2018ApJ...853..171G}, numerous massive (proto)stars and HII regions \citep{1995ApJ...449..663G,2011A&A...530L...9Q,2018ApJ...853..171G}. The burst of star formation suggests a wide spread gravitational instabilities among these cores. 

With the derived magnetic field estimations, we can characterize the significance between the magnetic field and gravity of the individual cores using the mass-to-flux ratio, $\lambda$, in units of critical value $1/(2\pi\sqrt{G})$ \citep{1978PASJ...30..671N,2004ApJ...600..279C}:\begin{equation}
    \lambda=\mu_\mathrm{H_2}m_\mathrm{H}\sqrt{\mu_0\pi G}\frac{N_\mathrm{H_2}}{B}\sim 7.6\times10^{-21}\frac{N_\mathrm{H_2}/\mathrm{cm}^{-2}}{B_\mathrm{3D}/\mu G}
\end{equation}
where $N_\mathrm{H_2}$ is the molecular hydrogen column density, and $B_\mathrm{3D}$ is the 3D magnetic field strength. Since we do not have the estimation of the line-of-sight component of the magnetic field strength, the 3D magnetic field strength $B_\mathrm{3D}$ can be estimated from the POS field based on the statistical relation. Assuming that the inclination angles between the underlying ordered 3D and POS magnetic fields are randomly distributed, \citet{2004ApJ...600..279C} proposed a statistical relation $B_\mathrm{3D}=4B^\mathrm{tot}_\mathrm{pos}/\pi$.

MHD simulations \citep[e.g.,][]{2011ApJ...742L...9C}{}{} show that the strongly magnetized cores with small mass-to-flux ratios suppress star formation while moderately magnetized cores are efficient in forming stars. In Sgr B2, these three dense cores undergoing starbursts are expected to have relatively weak magnetic fields with high mass-to-flux ratios. At a 0.2 pc scale, the calculated $\lambda$ values for Sgr B2(N), (M), and (S) are 17.7, 9.4, and 5.4, respectively. If we apply correction factor $Q_c=0.5$ to estimate $B^\mathrm{tot}_\mathrm{pos}$, the corresponding mass-to-flux ratios for Sgr B2(N), (M), and (S) are 7.4, 3.9, and 2.3, respectively. Even though the magnetic fields in the three cores are stronger than those in the Galactic disk, all three dense cores are magnetically supercritical ($\lambda>1$), meaning that gravity dominates the magnetic field. It is worth noting that Sgr B2(N), with its spiral-like magnetic field, has the highest mass-to-flux ratio. This is consistent with the notion that gravity is the dominant force that pulls the magnetic field lines toward the center of collapse, similar to that in IRAS 18089-1732 \citep[][]{2021ApJ...915L..10S}{}{} and G327.3 \citep[][]{2020ApJ...904..168B}{}{}. Conversely, Sgr B2(S) presents the lowest mass-to-flux ratio, exhibiting a nearly uniform distribution of magnetic fields aligned with the large-scale magnetic field orientation from the SOFIA data. This suggests that even though the magnetic field in this region can not resist gravitational collapse, it is still strong enough to maintain the orientation from its parental structure. At the cloud scale of 10 pc, the magnetic field morphology inferred from the SOFIA observations is nearly uniform within the three dense cores while undergoing significant variations outside these cores. The estimated mass-to-flux ratio of the region over 800$\sigma_I$ ranges from 8.5 ($Q_c=0.5$) to 20.3 ($Q_c=0.21$), suggesting that the large-scale magnetic field of Sgr B2 is also supercritical. We also analyzed the more extended structures surrounding Sgr B2 with an intensity cut of $I/\sigma_I>600$, which has a relatively lower density. Even though the derived density drops, the supercritical condition remains on a more extended structure of Sgr B2. In general, from the cloud scale down to the core scale, gravity always dominates the magnetic field in Sgr B2.

The relative importance between the magnetic field and turbulence of the individual cores can be characterized by Alfv$\mathrm{\acute{e}}$nic Mach number:
\begin{equation}
\mathcal{M}_A=\sigma_\mathrm{turb,3D}/\upsilon_\mathrm{A,3D}
\end{equation}
where $\sigma_\mathrm{turb,3D}=\sqrt{3}\sigma_\mathrm{turb}$ is an estimate for the 3D turbulent velocity dispersion, $\upsilon_\mathrm{A,3D}=B_\mathrm{3D}/\sqrt{\mu_0\rho}$ is the 3D Alfv$\mathrm{\acute{e}}$n velocity. The Alfv$\mathrm{\acute{e}}$nic Mach number for Sgr B2(N), (M), (S), and the Sgr B2 cloud are 4.9, 3.3, 3.1, and 6.1, respectively. If we apply correction factor $Q_c=0.5$ to estimate $B^\mathrm{tot}_\mathrm{pos}$, the corresponding Alfv$\mathrm{\acute{e}}$nic Mach number for Sgr B2(N), (M), (S), and the Sgr B2 cloud are 2.1, 2.0, 1.3, and 2.6, respectively. From the cloud scale to the core scale, Sgr B2 is in super-Alfv$\mathrm{\acute{e}}$nic states (weakly magnetized), meaning that the magnetic fields in different scale in Sgr B2 are not strong enough to resist turbulent compression \citep{2019FrASS...6....7K}.

The virial parameters are usually used to assess the stability of a core against gravitational collapse. Previous studies have made virial and dynamical analysis in Sgr B2 \citep[e.g,][they found over 70\% clumps in Sgr B2 are gravitationally bound with ongoing star formation]{2022ApJ...929...34M}. However, in their analyses, they assumed the magnetic field strength was negligible, while the magnetic ﬁeld could provide significant support to regulate the star formation. Following Appendix in \cite{2020ApJ...895..142L}, the virial parameters can be estimated with magnetic energy and kinetic energy considered:
\begin{equation}
    \alpha_\mathrm{k+B}=\frac{M_\mathrm{k+B}}{M_\mathrm{gas}}
\end{equation}
where $M_\mathrm{k+B}=\sqrt{M_B^2+(M_\mathrm{k}/2)^2}+M_\mathrm{k}/2$ is the critical virial mass, and $M_\mathrm{gas}$ is the gas mass of the source. At 0.2 pc scale, assuming a uniform density profile, all three dense cores are found to be sub-virial or trans-virial ($\alpha_\mathrm{k+B}\lesssim1$), especially Sgr B2(N) and (M) having virial parameters much less than 1. If a centrally peaked density profile ($\rho\propto r^{-2}$) is assumed, the derived viral parameters are even smaller (see Table \ref{tab:adfpara}). At a 10 pc scale, the derived virial parameter for Sgr B2 complex is about 0.20 (uniform density profile) and 0.13 (centrally peaked density profile). If we apply correction factor $Q_c=0.5$, all the sources are still sub-virial or trans-virial. In general, from 10 pc scale cloud to 0.2 pc scale cores, these structures are all unstable against gravitational collapse, which is consistent with the ongoing starburst in Sgr B2. 

\subsection{Comparison with other starburst regions}
There are several other star-forming regions undergoing mini-starburst \citep[with star formation rate density of 10-100 $M_\odot\mathrm{\ yr^{-1}\ kpc^{-2}}$,][]{2003ApJ...582..277M, 2012sf2a.conf...45M, 2013ApJ...778...96W}{}{} in the Galaxy, such as W43-MM1 \citep[with star formation rate about $0.006\ \mathrm{M_\odot yr^{-1}}$ enclosed in 8 pc$^3$ from][]{2014A&A...570A..15L}{}{}. In W43-MM1, an ordered polarization pattern was observed by \citet{2006ApJ...639..965C} using BIMA (Berkeley-Illinois-Maryland Association). At the clump scale, the estimated mass-to-flux ratio was 1.9 \citep{2010A&A...519A..35C}. Later, \cite{2014ApJ...783L..31S} used the SMA 345 GHz polarization observations with higher resolution ($\sim 2\arcsec$) to obtain a pinched morphology of magnetic fields in W43-MM1 with a mass-to-flux ratio of 3.5. Recent ALMA observations at 1 mm (Band 6) from \cite{2016ApJ...825L..15C}
resolved W43-MM1 down to 0.01 pc scale. They found all fragments inside it are super-critical. Similar to Sgr B2, the magnetically super-critical condition is maintained on smaller spatial scales in W43-MM1.

NGC 6334 is another massive star-forming complex undergoing starburst \citep[][]{2013ApJ...778...96W,2017A&A...607A..86R}{}{}. The main molecular gas structure in this complex is a long ﬁlamentary cloud consisting of six massive star-forming clumps (NGC6334 I-V and I(N), ``N'' denotes north). \cite{2015Natur.520..518L} carried out a multi-scale study on the magnetic field in NGC6334 from a cloud (10 pc), clump (1 pc) to core (0.1 pc) scale. Assuming a force balance between gravity, the magnetic pressure and magnetic tension, they found that all these structures are magnetically super-critical. The estimated mass-to-flux ratios normalized to critical value are $1.1\pm0.24$, $2.4\pm0.74$, $2.2\pm0.54$, $1.7\pm0.32$ for the whole cloud, clump I/I(N), core I and core I(N), respectively. Meanwhile, \citet{2023ApJ...945..160L} also investigated the multi-scale properties of NGC 6334 using polarization data from various instruments, including Planck ($\sim$25 pc), JCMT ($\sim$1 pc), and ALMA ($\sim$0.1 pc). They derived the normalized mass-to-flux ratio ($\lambda_\mathrm{KTH}$) from the KTH method as a function of column densities measured from different instruments. They found that most low column density bins ($N_\mathrm{H_2}<5\times10^{22}\mathrm{cm^{-2}}$) in the Planck observations, which correspond to regions outside the NGC6334 filament, exhibit subcritical ($\lambda_\mathrm{KTH}<1$) condition. On the other hand, the highest density bin ($N_\mathrm{H_2}\sim1\times10^{23}\mathrm{cm^{-2}}$), covering the 10-pc-scale NGC 6334 filament, indicates $\lambda_\mathrm{KTH}>1$, consistent with the supercritical estimation by \citep{2015Natur.520..518L}. For JCMT and ALMA observations, $\lambda_\mathrm{KTH}$ increases with increasing $N_{H_2}$, and transitions from subcritical to supercritical state. However, it is worth noting that the uncertainties of $\lambda_\mathrm{KTH}$ from the KTH method have not been well constrained, and the uncertainties on the column density of the Planck and JCMT data are also unclear. If we focus on the highest density structures ($\sim10^{23}~\mathrm{cm^{-2}}$) observed by different instruments, the supercritical condition remains from the cloud-scale to core-scale in NGC 6334.

Based on the multi-scale studies of mini-starburst regions of W43, NGC 6334, and Sgr B2, it appears that maintaining supercritical conditions across different scales may be necessary in order to initiate intense star formation activities inside them. However, it is worth noting that the relatively weak magnetic field may not be the sole determinant of mini starburst in these regions. For instance, all these mini-starburst regions harbor numerous HII regions \citep[][]{1995ADIL...DM...02M, 2008ApJS..177..515L,1999AJ....117.1392B,2008hsf2.book..456P}{}{}, whose expansion can compress surrounding neutral gas into denser regions, thus also fostering star formation activities \citep[][]{2005A&A...433..565D,2014prpl.conf..149T}{}{}. Further investigations, particularly comparative analyses with non-starburst regions, will provide further insights into the driving force behind mini-starbursts in these regions.

\section{Summary}\label{sec:summary}
With the high-resolution SMA observations, we present the magnetic field morphology of three dense cores in Sgr B2. Applying the angular dispersion function analysis, we derived magnetic field strengths of 4.3-10.0, 6.2-14.7, and 1.9-4.5 mG for Sgr B2(N), (M), and (S), respectively, with a correction factors of 0.21 and 0.5. Based on different methods, we found that magnetic fields in these three cores are all dominated by gravity and the cores are all subvirial, which is consistent with the mini starburst in Sgr B2. Additionally, the parsec-scale magnetic field in Sgr B2 inferred by the SOFIA data is also found to be dominated by gravity, indicating that the supercritical conditions are preserved from large-scale ($\sim$10 pc) to small-scale ($\sim$0.2 pc) in Sgr B2.

\begin{acknowledgements}
    We thank Dr. Dylan Par$\mathrm{\acute{e}}$ and the SOFIA/HAWC+ Far-Infrared Polarimetric Large Area CMZ Exploration (FIREPLACE) team for sharing their DR2 datasets in the CMZ. X. P. is supported by the Smithsonian Astrophysical Observatory (SAO) Predoctoral Fellowship Program. Q. Z. acknowledges the support of National Science Foundation under award No. 2206512. The authors wish to recognize and acknowledge the very significant cultural role and reverence that the summit of Maunakea has always had within the indigenous Hawaiian community. We are most fortunate to have had the opportunity to conduct observations from this mountain. This work used observations made with the NASA/DLR Stratospheric Observatory for Infrared Astronomy (SOFIA). SOFIA was jointly operated by the Universities Space Research Association, Inc. (USRA), under NASA contract NNA17BF53C, and the Deutsches SOFIA Institut (DSI) under DLR contract 50 OK 2002 to the University of Stuttgart.
\end{acknowledgements}

\vspace{5mm}
% \facilities{}

%% Similar to \facility{}, there is the optional \software command to allow 
%% authors a place to specify which programs were used during the creation of 
%% the manuscript. Authors should list each code and include either a
%% citation or url to the code inside ()s when available.

% \software{astropy \citep{2013A&A...558A..33A,2018AJ....156..123A},  
%           Cloudy \citep{2013RMxAA..49..137F}, 
%           Source Extractor \citep{1996A&AS..117..393B}
%           }

%% Appendix material should be preceded with a single \appendix command.
%% There should be a \section command for each appendix. Mark appendix
%% subsections with the same markup you use in the main body of the paper.

%% Each Appendix (indicated with \section) will be lettered A, B, C, etc.
%% The equation counter will reset when it encounters the \appendix
%% command and will number appendix equations (A1), (A2), etc. The
%% Figure and Table counter will not reset.
\clearpage
\appendix
\section{Effect of Errors on Instrumental polarization Calibration}\label{appen:pol_leakage}
\begin{figure}[!ht]
    \centering
    \gridline{\fig{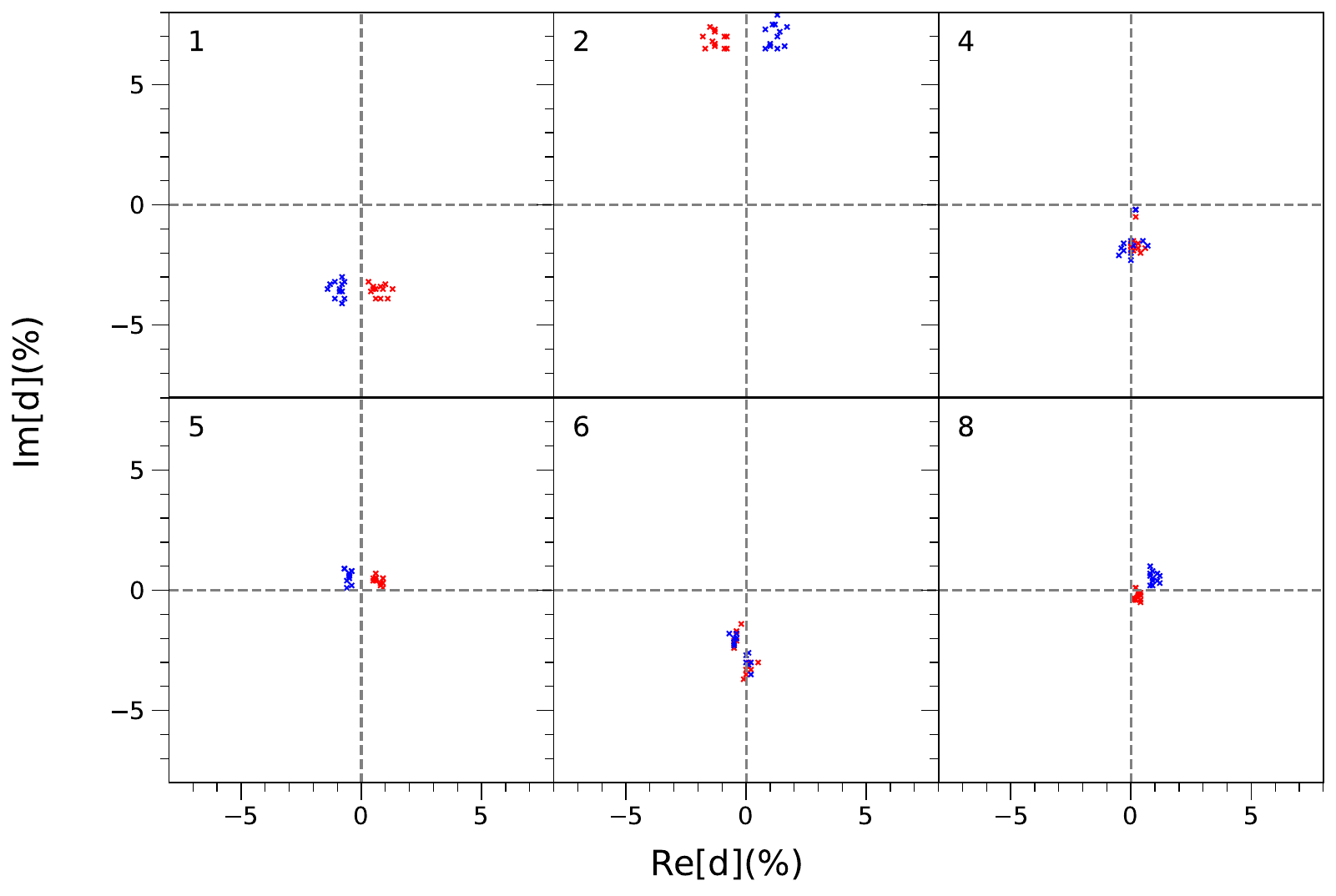}{0.7\textwidth}{}}
    \gridline{
    \fig{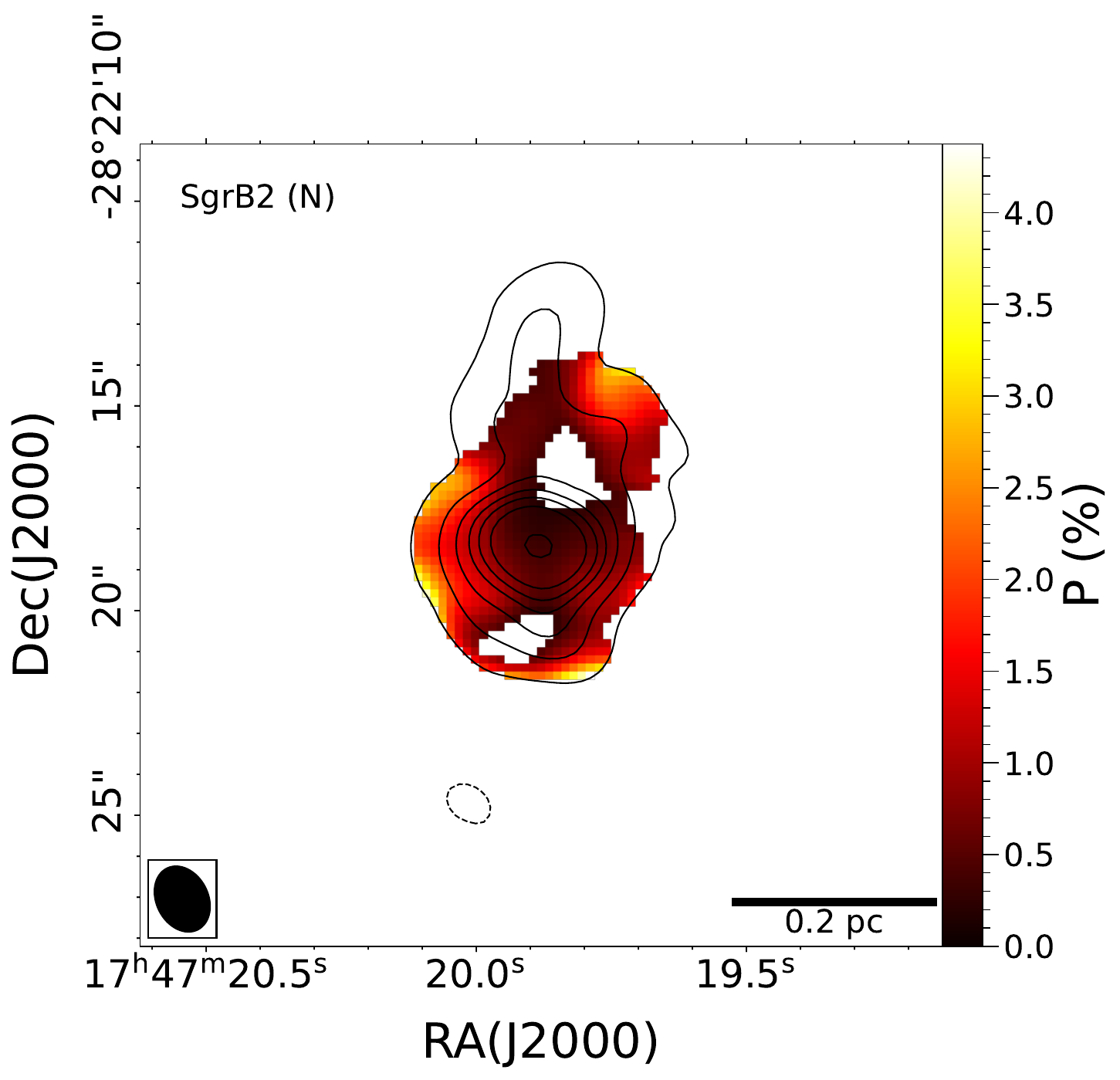}{0.3\textwidth}{}
    \fig{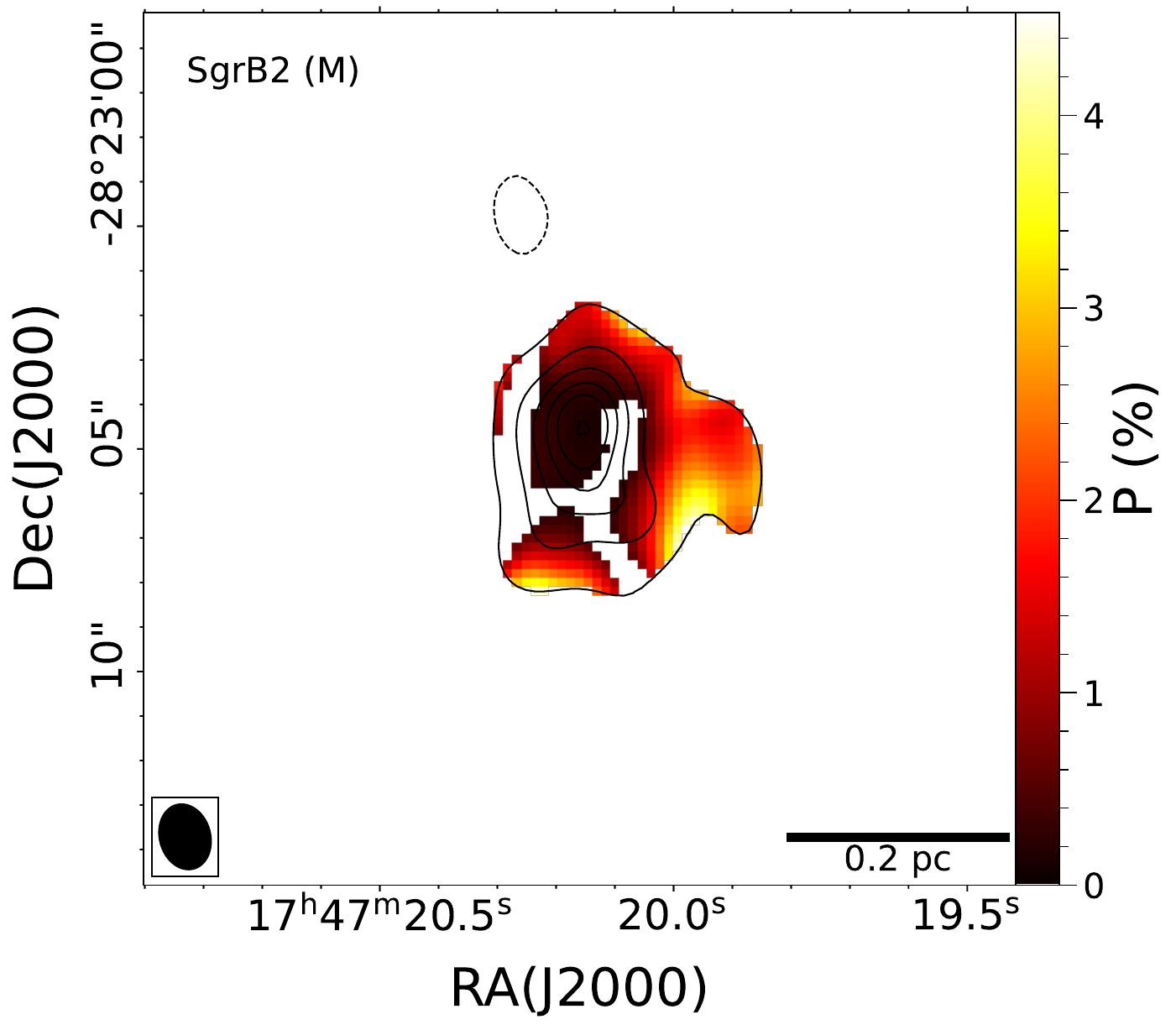}{0.3\textwidth}{}
    \fig{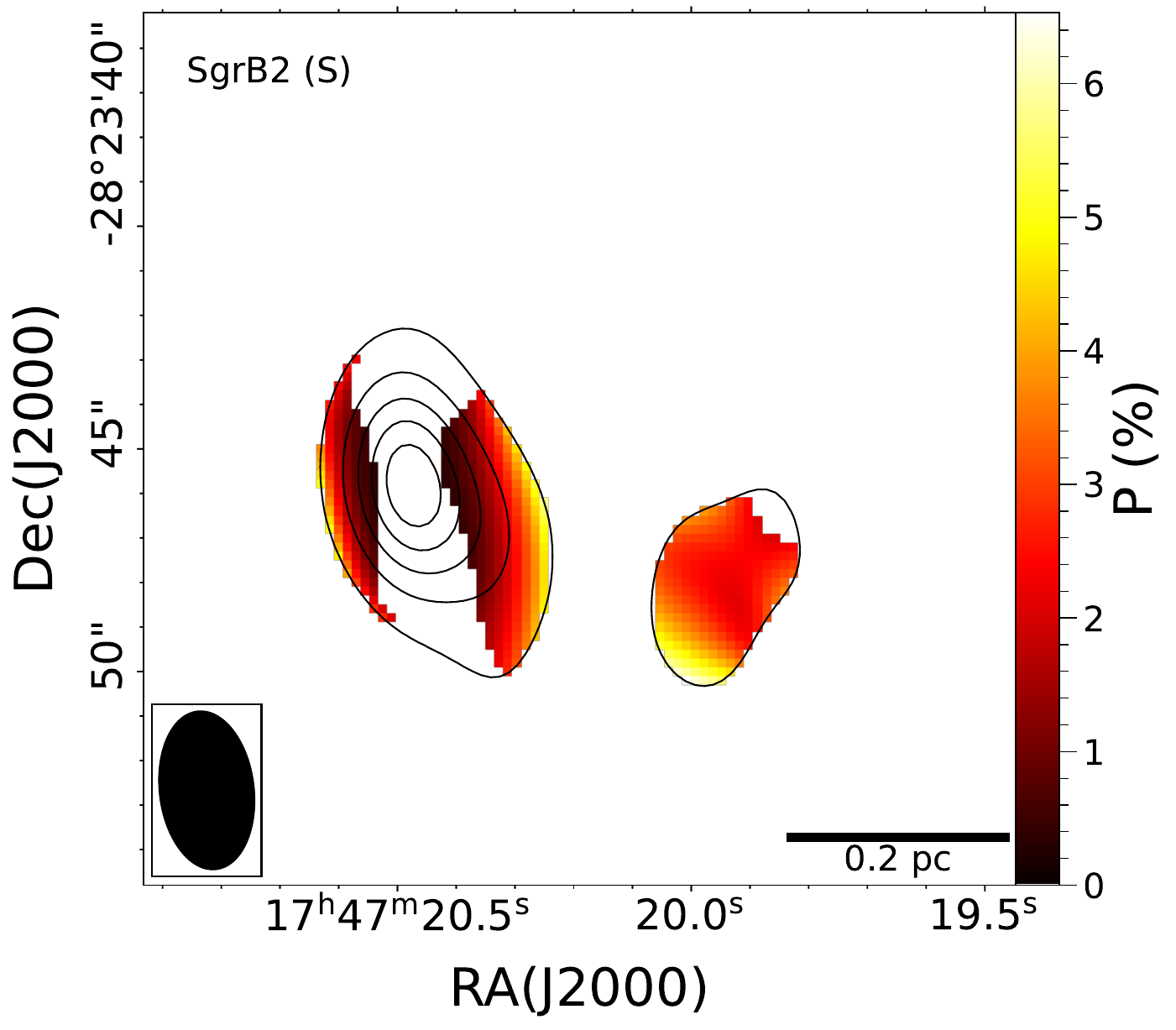}{0.3\textwidth}{}
    }
    \caption{Top: Leakage terms measured in the 345 GHz band for each antenna from all six SMA tracks. Antenna number is labeled at the upper left corner of each panel. Leakages are plotted in the complex plane, with equal linear scales in the real ($\mathrm{Re}[d]$) and imaginary ($\mathrm{Im}[d]$) axes. The red and blue crosses represent $d_L$ and $d_R$, respectively. Antennas 3 and 7 are not available during the observations. Bottom: Dust polarization percentage maps for Sgr B2 from the SMA observations. The polarization percentage is shown in color scales. Black solid contour represents Stoke I emission. Contour levels are the same as Fig. \ref{fig:SgrB2_mag}.}
    \label{fig:leakage}
\end{figure}

The imperfect splitting of the incoming radiation into the two orthogonal polarized components results in the mixing of one component into the other. This is referred to as the instrumental polarization or leakage which is an antenna-based quantity that also depends on the frequency. The leakage terms are determined by observing a polarization calibrator over a wide range of parallactic angle. The antenna based leakage terms are fixed in the frame of reference of the earth while the measured calibrator polarization angle rotates as a function of the parallactic angle. This allows us to simultaneously solve for both the leakage terms and the calibrator linear polarization. The leakage terms for different antennas of our SMA data during different tracks of the Sgr B2 observations are plotted in Fig. \ref{fig:leakage}. The SMA polarimetry system has very stable instrumental polarization as a consequence of the optical design \citep{2006PhDT........32M}. As a first estimate, we make the assumption that the dispersion of the leakages on different days is a good representation of the uncertainty in the leakages. We found that the average leakage variations for different antennas range from 0.2\% to 0.6\%. This estimate is an upper bound as there may be some small time variations. These variations could be a result of factors such as i) inadequate parallactic angle coverage of the polarization calibrator, ii) errors in the data, iii) real changes that affect the optics, or iv) second order effects due to strong source polarization. In particular, we can see that for antenna 4, there is one anomalous pair of points which was taken on June 10, 2022. In addition, antenna 6 shows two different clusters of measurements. After removing these anomalous points, we calculate an average leakage uncertainty of $\sim$ 0.23\% across all antennas.  

We conducted simulations to determine the effect of the uncertainty in the leakage terms on the polarization data of the science target. We created a customized version of the UVGEN program from the MIRIAD data reduction package which mimicked the SMA polarimetry system. The test source was chosen to be an unpolarized point source at phase center with the same declination as the target source (Sgr~B2). The dataset had an hour angle coverage similar to one of the actual SMA observations. The leakages were generated with a random distribution around a mean zero and with the standard deviation equal to the uncertainty of 0.23\% as determined above. The computed correlations in the simulated dataset were affected by uncalibrated leakages. Not correcting for this error resulted in an unpolarized source being corrupted by $\sim$0.09\% polarization on average, i.e. an unpolarized source at phase center appears to show a spurious polarization of 0.09\% of the Stokes I emission. Simulations were made using different values of the leakage uncertainty and we found that uncorrected and random errors led to a spurious polarization in the final image which is lower by a factor  of $\sim$3, i.e., a leakage uncertainty of 0.23\% leads to a spurious image polarization of 0.09\%. This reduction is likely due to a combination of the random nature of the errors along with the parallactic angle coverage of the target. Fig. \ref{fig:leakage} also shows the polarization percentage of the three sources detected by the SMA, most of which are above the 2$\sigma$ level of 0.18\%.

\section{Determination of the Turbulence}\label{appen:turbulence}
\begin{figure}[!ht]
    \centering
    \gridline{
    \fig{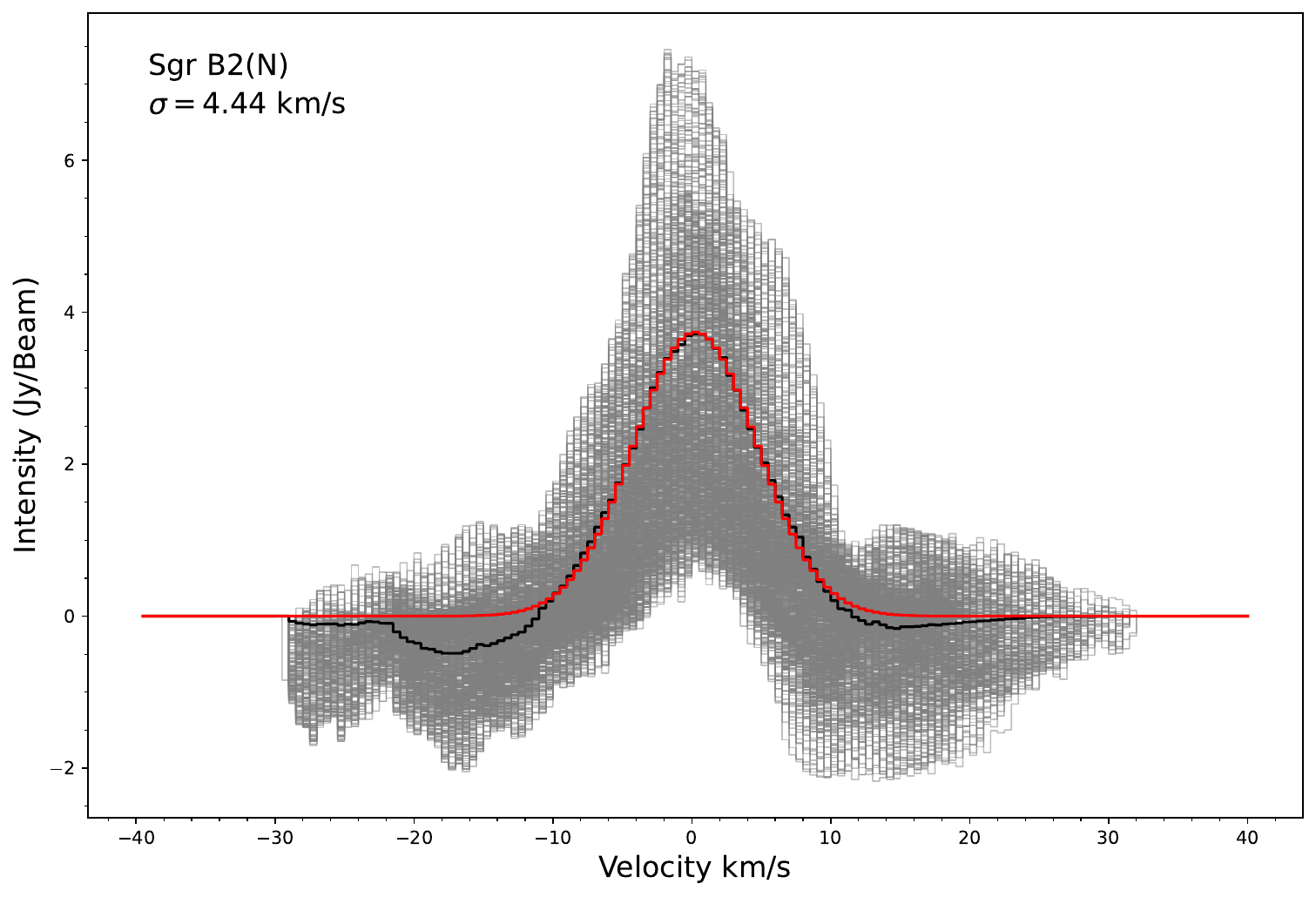}{0.33\textwidth}{}
    \fig{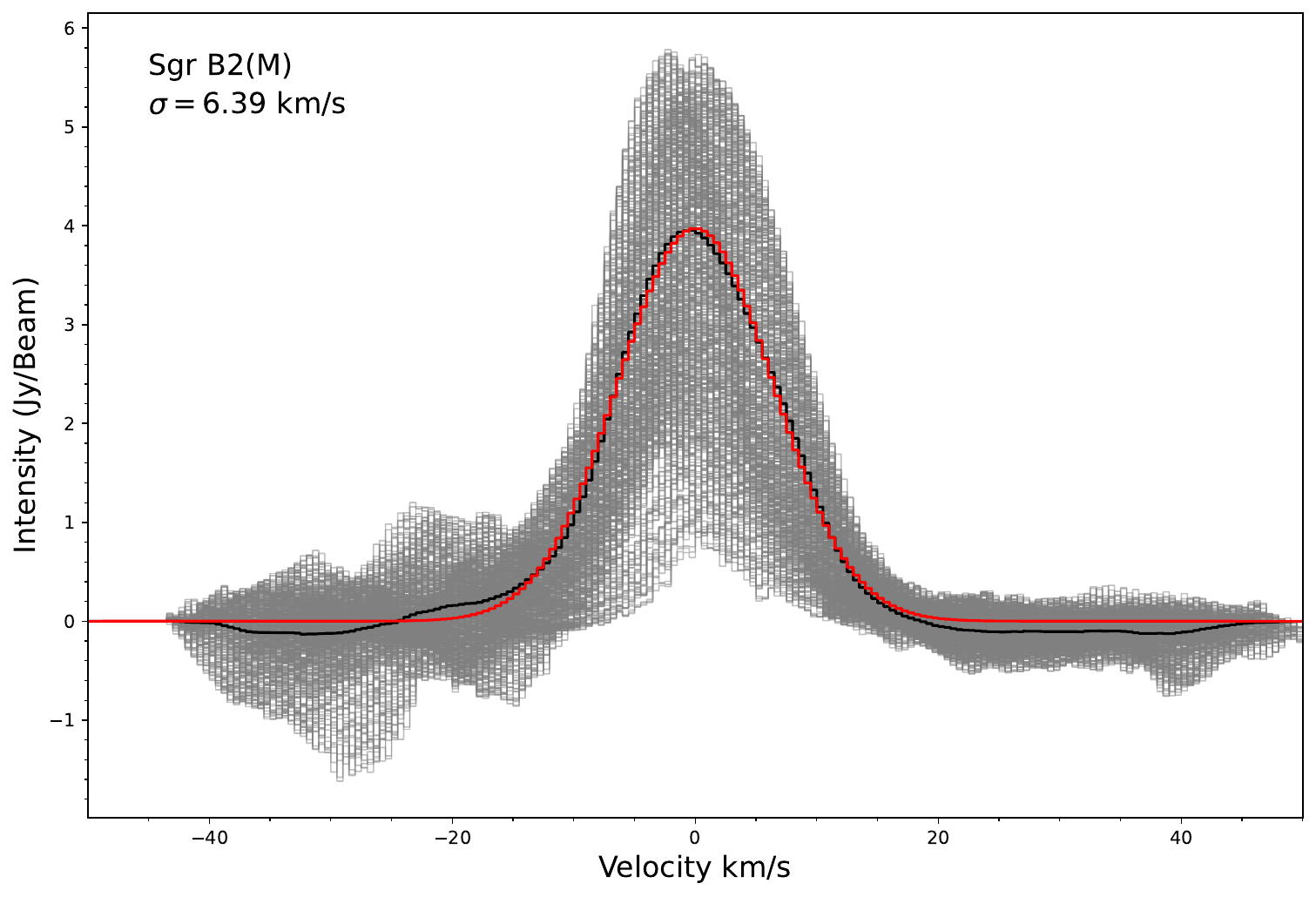}{0.33\textwidth}{}
    \fig{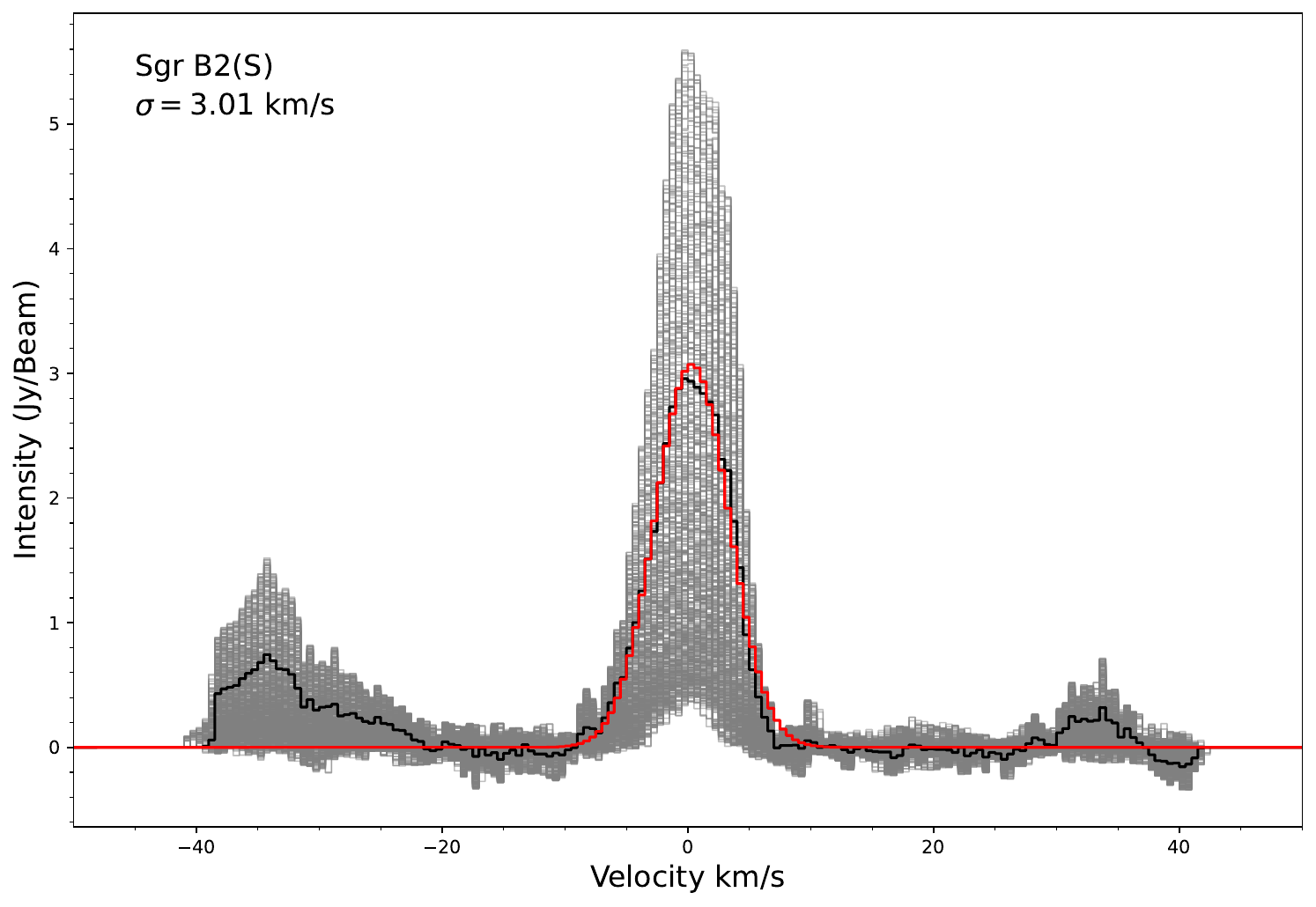}{0.33\textwidth}{}
          }
    \caption{$\mathrm{CH_3OCHO}$ spectra of the three cores corrected by the local velocity (corresponding to the local value from intensity-weight velocity maps). The gray lines represent all corrected $\mathrm{CH_3OCHO}$ spectra in each core over 3$\sigma$ contours of Stoke I emission. The black line represents the averaged spectrum weighted by the integrated intensity of each pixel. The red line shows the Gaussian fitting result. The fitted velocity dispersion from turbulence is labeled upper left.}
    \label{fig:turb_CH3OCHO}
\end{figure}

The bulk motion within cores would increase the velocity dispersion, which leads to an overestimate of the turbulent motions. To separate the small-scale turbulent fluctuations from the large-scale bulk motion, we shifted the spectra of each pixel to the local velocity (intensity-weighted mean velocity) and averaged all the spectra within the dense core. Then, we fitted the average spectra with a Gaussian function to the averaged spectra to determine the velocity dispersion for each core. Fig. \ref{fig:turb_CH3OCHO} illustrates the fitting results from the three cores. The contribution from the thermal motion to the velocity dispersion is negligible here.

\section{Angular Dispersion Function Analysis}\label{appen:adfmethod}
\citet{2016ApJ...820...38H} derived the angular dispersion solutions for polarimetric images obtained from a single-dish telescope and an interferometer.

The single-dish solution can be expressed as:
\begin{equation}
    1-\langle\cos[\triangle\phi(\ell)]\rangle=\sum^\infty_{j=1}a_{2j}\ell^{2j}+\left[\frac{1}{1+N_1\langle B^2_0\rangle/\langle B^2_t\rangle}\right]\left[1-e^{-\ell^2/2(\delta^2+2W^2_1)}\right]
    \label{equ:singledish}
\end{equation}
where $\Delta\phi(\ell)$ is the angular dispersion of two polarization segments as a function of the distance, $\ell$, between them, $\delta$ is the turbulence correlation length, $\langle B^2_0\rangle/\langle B^2_t\rangle$ is the ratio of between the ordered magnetic field and the turbulent magnetic field, and $N_1$ is the number of turbulent cells probed by the telescope beam:
\begin{equation}
    N_1=\frac{(\delta^2+2W_1^2)\Delta^\prime}{\sqrt{2\pi}\delta^3}
\end{equation}
$\Delta^\prime$ is the effective thickness of the source, which can be estimated from the auto-correlation function of polarized flux \citep[see][]{2009ApJ...706.1504H}{}{}. We can derive the dispersion function of polarization angles (left hand of Equation \ref{equ:singledish}), which can be fitted with a model (right hand of Equation \ref{equ:singledish}) to obtain the estimation of $\delta$, $\langle B^2_0\rangle/\langle B^2_t\rangle$. The signal-integrated turbulence auto-correlation function $b^2(\ell)$ is:
\begin{equation}
    b^2(\ell)=\left[\frac{1}{1+N_1\langle B^2_0\rangle/\langle B^2_t\rangle}\right]e^{-\ell^2/2(\delta^2+2W^2_1)}
\end{equation}

The interferometer solution can be expressed as:
\begin{equation}
    1-\langle\cos[\triangle\phi(\ell)]\rangle=\sum^\infty_{j=1}a_{2j}\ell^{2j}+\frac{1}{1+N\langle B^2_0\rangle/\langle B^2_t\rangle}-b^2(\ell)
\end{equation}
where \emph{N} is the number of turbulent cells probed by the telescope beam, $b^2(\ell)$ is the local turbulent component for the interferometer solution. Taking $W_1$ and $W_2$ as the standard deviation of the two Gaussian profiles of the synthesized beam ($\sqrt{\mathrm{FWHM_{maj}\times FWHM_{min}}}/\sqrt{8\ln2}$) and the large-scale filtering effect, we can express \emph{N} and $b^2(\ell)$ as:
\begin{equation}
    N_1=\frac{(\delta^2+2W_1^2)\Delta^\prime}{\sqrt{2\pi}\delta^3}
\end{equation}
\begin{equation}
    N_2=\frac{(\delta^2+2W_2^2)\Delta^\prime}{\sqrt{2\pi}\delta^3}
\end{equation}
\begin{equation}
    N_{12}=\frac{(\delta^2+W_1^2+W_2^2)\Delta^\prime}{\sqrt{2\pi}\delta^3}
\end{equation}
\begin{equation}
    N=\left(\frac{1}{N_1}+\frac{1}{N_2}-\frac{2}{N_{12}}\right)^{-1}
\end{equation}
\begin{equation}
    b^2(\ell)=\left[\frac{N}{1+N\langle B^2_0\rangle/\langle B^2_t\rangle}\right]\times\left[\frac{1}{N_1}e^{-\ell^2/2(\delta^2+2W_1^2)}+\frac{1}{N_2}e^{-\ell^2/2(\delta^2+2W_2^2)}-\frac{2}{N_{12}}e^{-\ell^2/2(\delta^2+2W_1^2+2W_2^2)}\right]
\end{equation}
Fig. \ref{fig:ADF_fit} shows the Sgr B2 polarization angular dispersion functions and fitting results for SMA and SOFIA data. The best ﬁt is obtained via $\chi^2$ minimization.

% \section{KTH Method}\label{appen:kth}

%% For this sample we use BibTeX plus aasjournals.bst to generate the
%% the bibliography. The sample631.bib file was populated from ADS. To
%% get the citations to show in the compiled file do the following:
%%
%% pdflatex sample631.tex
%% bibtext sample631
%% pdflatex sample631.tex
%% pdflatex sample631.tex
\clearpage
\bibliography{SMA_SgrB2_ref}{}

\begin{thebibliography}{}
\expandafter\ifx\csname natexlab\endcsname\relax\def\natexlab#1{#1}\fi
\providecommand{\url}[1]{\href{#1}{#1}}
\providecommand{\dodoi}[1]{doi:~\href{http://doi.org/#1}{\nolinkurl{#1}}}
\providecommand{\doeprint}[1]{\href{http://ascl.net/#1}{\nolinkurl{http://ascl.net/#1}}}
\providecommand{\doarXiv}[1]{\href{https://arxiv.org/abs/#1}{\nolinkurl{https://arxiv.org/abs/#1}}}

\bibitem[{{Andersson} {et~al.}(2015){Andersson}, {Lazarian}, \&
  {Vaillancourt}}]{2015ARA&A..53..501A}
{Andersson}, B.~G., {Lazarian}, A., \& {Vaillancourt}, J.~E. 2015, \araa, 53,
  501, \dodoi{10.1146/annurev-astro-082214-122414}

\bibitem[{{Barnes} {et~al.}(2017){Barnes}, {Longmore}, {Battersby}, {Bally},
  {Kruijssen}, {Henshaw}, \& {Walker}}]{2017MNRAS.469.2263B}
{Barnes}, A.~T., {Longmore}, S.~N., {Battersby}, C., {et~al.} 2017, \mnras,
  469, 2263, \dodoi{10.1093/mnras/stx941}

\bibitem[{{Beltr{\'a}n} {et~al.}(2019){Beltr{\'a}n}, {Padovani}, {Girart},
  {Galli}, {Cesaroni}, {Paladino}, {Anglada}, {Estalella}, {Osorio}, {Rao},
  {S{\'a}nchez-Monge}, \& {Zhang}}]{2019A&A...630A..54B}
{Beltr{\'a}n}, M.~T., {Padovani}, M., {Girart}, J.~M., {et~al.} 2019, \aap,
  630, A54, \dodoi{10.1051/0004-6361/201935701}

\bibitem[{{Beuther} {et~al.}(2020){Beuther}, {Soler}, {Linz}, {Henning},
  {Gieser}, {Kuiper}, {Vlemmings}, {Hennebelle}, {Feng}, {Smith}, \&
  {Ahmadi}}]{2020ApJ...904..168B}
{Beuther}, H., {Soler}, J.~D., {Linz}, H., {et~al.} 2020, \apj, 904, 168,
  \dodoi{10.3847/1538-4357/abc019}

\bibitem[{{Blum} {et~al.}(1999){Blum}, {Damineli}, \&
  {Conti}}]{1999AJ....117.1392B}
{Blum}, R.~D., {Damineli}, A., \& {Conti}, P.~S. 1999, \aj, 117, 1392,
  \dodoi{10.1086/300791}

\bibitem[{{Bonfand} {et~al.}(2017){Bonfand}, {Belloche}, {Menten}, {Garrod}, \&
  {M{\"u}ller}}]{2017A&A...604A..60B}
{Bonfand}, M., {Belloche}, A., {Menten}, K.~M., {Garrod}, R.~T., \&
  {M{\"u}ller}, H.~S.~P. 2017, \aap, 604, A60,
  \dodoi{10.1051/0004-6361/201730648}

\bibitem[{{Budaiev} {et~al.}(2024){Budaiev}, {Ginsburg}, {Jeff}, {Goddi},
  {Meng}, {S{\'a}nchez-Monge}, {Schilke}, {Schmiedeke}, \&
  {Yoo}}]{2024ApJ...961....4B}
{Budaiev}, N., {Ginsburg}, A., {Jeff}, D., {et~al.} 2024, \apj, 961, 4,
  \dodoi{10.3847/1538-4357/ad0383}

\bibitem[{{Busch} {et~al.}(2023){Busch}, {Belloche}, {Garrod}, {Mueller}, \&
  {Menten}}]{2023arXiv231011339B}
{Busch}, L.~A., {Belloche}, A., {Garrod}, R.~T., {Mueller}, H. S.~P., \&
  {Menten}, K.~M. 2023, arXiv e-prints, arXiv:2310.11339,
  \dodoi{10.48550/arXiv.2310.11339}

\bibitem[{{Butterfield} {et~al.}(2023){Butterfield}, {Chuss}, {Guerra},
  {Morris}, {Pare}, {Wollack}, {Dowell}, {Hankins}, {Karpovich}, {Siah},
  {Staguhn}, \& {Zweibel}}]{2023arXiv230601681B}
{Butterfield}, N.~O., {Chuss}, D.~T., {Guerra}, J.~A., {et~al.} 2023, arXiv
  e-prints, arXiv:2306.01681, \dodoi{10.48550/arXiv.2306.01681}

\bibitem[{{Caswell} {et~al.}(2010){Caswell}, {Fuller}, {Green}, {Avison},
  {Breen}, {Brooks}, {Burton}, {Chrysostomou}, {Cox}, {Diamond}, {Ellingsen},
  {Gray}, {Hoare}, {Masheder}, {McClure-Griffiths}, {Pestalozzi}, {Phillips},
  {Quinn}, {Thompson}, {Voronkov}, {Walsh}, {Ward-Thompson}, {Wong-McSweeney},
  {Yates}, \& {Cohen}}]{2010MNRAS.404.1029C}
{Caswell}, J.~L., {Fuller}, G.~A., {Green}, J.~A., {et~al.} 2010, \mnras, 404,
  1029, \dodoi{10.1111/j.1365-2966.2010.16339.x}

\bibitem[{{Chandrasekhar} \& {Fermi}(1953)}]{1953ApJ...118..113C}
{Chandrasekhar}, S., \& {Fermi}, E. 1953, \apj, 118, 113,
  \dodoi{10.1086/145731}

\bibitem[{{Commer{\c{c}}on} {et~al.}(2011){Commer{\c{c}}on}, {Hennebelle}, \&
  {Henning}}]{2011ApJ...742L...9C}
{Commer{\c{c}}on}, B., {Hennebelle}, P., \& {Henning}, T. 2011, \apjl, 742, L9,
  \dodoi{10.1088/2041-8205/742/1/L9}

\bibitem[{{Cortes} \& {Crutcher}(2006)}]{2006ApJ...639..965C}
{Cortes}, P., \& {Crutcher}, R.~M. 2006, \apj, 639, 965, \dodoi{10.1086/498971}

\bibitem[{{Cortes} {et~al.}(2010){Cortes}, {Parra}, {Cortes}, \&
  {Hardy}}]{2010A&A...519A..35C}
{Cortes}, P.~C., {Parra}, R., {Cortes}, J.~R., \& {Hardy}, E. 2010, \aap, 519,
  A35, \dodoi{10.1051/0004-6361/200811137}

\bibitem[{{Cortes} {et~al.}(2016){Cortes}, {Girart}, {Hull}, {Sridharan},
  {Louvet}, {Plambeck}, {Li}, {Crutcher}, \& {Lai}}]{2016ApJ...825L..15C}
{Cortes}, P.~C., {Girart}, J.~M., {Hull}, C. L.~H., {et~al.} 2016, \apjl, 825,
  L15, \dodoi{10.3847/2041-8205/825/1/L15}

\bibitem[{{Cort{\'e}s} {et~al.}(2021){Cort{\'e}s}, {Sanhueza}, {Houde},
  {Mart{\'\i}n}, {Hull}, {Girart}, {Zhang}, {Fernandez-Lopez}, {Zapata},
  {Stephens}, {Li}, {Wu}, {Olguin}, {Lu}, {Guzm{\'a}n}, \&
  {Nakamura}}]{2021ApJ...923..204C}
{Cort{\'e}s}, P.~C., {Sanhueza}, P., {Houde}, M., {et~al.} 2021, \apj, 923,
  204, \dodoi{10.3847/1538-4357/ac28a1}

\bibitem[{{Crutcher} {et~al.}(2004){Crutcher}, {Nutter}, {Ward-Thompson}, \&
  {Kirk}}]{2004ApJ...600..279C}
{Crutcher}, R.~M., {Nutter}, D.~J., {Ward-Thompson}, D., \& {Kirk}, J.~M. 2004,
  \apj, 600, 279, \dodoi{10.1086/379705}

\bibitem[{{Davis}(1951)}]{1951PhRv...81..890D}
{Davis}, L. 1951, Physical Review, 81, 890, \dodoi{10.1103/PhysRev.81.890.2}

\bibitem[{{Deharveng} {et~al.}(2005){Deharveng}, {Zavagno}, \&
  {Caplan}}]{2005A&A...433..565D}
{Deharveng}, L., {Zavagno}, A., \& {Caplan}, J. 2005, \aap, 433, 565,
  \dodoi{10.1051/0004-6361:20041946}

\bibitem[{{Dotson} {et~al.}(2010){Dotson}, {Vaillancourt}, {Kirby}, {Dowell},
  {Hildebrand}, \& {Davidson}}]{2010ApJS..186..406D}
{Dotson}, J.~L., {Vaillancourt}, J.~E., {Kirby}, L., {et~al.} 2010, \apjs, 186,
  406, \dodoi{10.1088/0067-0049/186/2/406}

\bibitem[{{Dowell}(1997)}]{1997ApJ...487..237D}
{Dowell}, C.~D. 1997, \apj, 487, 237, \dodoi{10.1086/304577}

\bibitem[{{Etxaluze} {et~al.}(2013){Etxaluze}, {Goicoechea}, {Cernicharo},
  {Polehampton}, {Noriega-Crespo}, {Molinari}, {Swinyard}, {Wu}, \&
  {Bally}}]{2013A&A...556A.137E}
{Etxaluze}, M., {Goicoechea}, J.~R., {Cernicharo}, J., {et~al.} 2013, \aap,
  556, A137, \dodoi{10.1051/0004-6361/201321258}

\bibitem[{{Falceta-Gon{\c{c}}alves} {et~al.}(2008){Falceta-Gon{\c{c}}alves},
  {Lazarian}, \& {Kowal}}]{2008ApJ...679..537F}
{Falceta-Gon{\c{c}}alves}, D., {Lazarian}, A., \& {Kowal}, G. 2008, \apj, 679,
  537, \dodoi{10.1086/587479}

\bibitem[{{Federrath} {et~al.}(2010){Federrath}, {Roman-Duval}, {Klessen},
  {Schmidt}, \& {Mac Low}}]{2010A&A...512A..81F}
{Federrath}, C., {Roman-Duval}, J., {Klessen}, R.~S., {Schmidt}, W., \& {Mac
  Low}, M.~M. 2010, \aap, 512, A81, \dodoi{10.1051/0004-6361/200912437}

\bibitem[{{Gaume} {et~al.}(1995){Gaume}, {Claussen}, {de Pree}, {Goss}, \&
  {Mehringer}}]{1995ApJ...449..663G}
{Gaume}, R.~A., {Claussen}, M.~J., {de Pree}, C.~G., {Goss}, W.~M., \&
  {Mehringer}, D.~M. 1995, \apj, 449, 663, \dodoi{10.1086/176087}

\bibitem[{{Ginsburg} {et~al.}(2018){Ginsburg}, {Bally}, {Barnes}, {Bastian},
  {Battersby}, {Beuther}, {Brogan}, {Contreras}, {Corby}, {Darling}, {De Pree},
  {Galv{\'a}n-Madrid}, {Garay}, {Henshaw}, {Hunter}, {Kruijssen}, {Longmore},
  {Lu}, {Meng}, {Mills}, {Ott}, {Pineda}, {S{\'a}nchez-Monge}, {Schilke},
  {Schmiedeke}, {Walker}, \& {Wilner}}]{2018ApJ...853..171G}
{Ginsburg}, A., {Bally}, J., {Barnes}, A., {et~al.} 2018, \apj, 853, 171,
  \dodoi{10.3847/1538-4357/aaa6d4}

\bibitem[{{Gordon} {et~al.}(1993){Gordon}, {Berkermann}, {Mezger}, {Zylka},
  {Haslam}, {Kreysa}, {Sievers}, \& {Lemke}}]{1993A&A...280..208G}
{Gordon}, M.~A., {Berkermann}, U., {Mezger}, P.~G., {et~al.} 1993, \aap, 280,
  208

\bibitem[{{Harper} {et~al.}(2018){Harper}, {Runyan}, {Dowell}, {Wirth},
  {Amato}, {Ames}, {Amiri}, {Banks}, {Bartels}, {Benford}, {Berthoud},
  {Buchanan}, {Casey}, {Chapman}, {Chuss}, {Cook}, {Derro}, {Dotson}, {Evans},
  {Fixsen}, {Gatley}, {Guerra}, {Halpern}, {Hamilton}, {Hamlin}, {Hansen},
  {Heimsath}, {Hermida}, {Hilton}, {Hirsch}, {Hollister}, {Hostetter}, {Irwin},
  {Jhabvala}, {Jhabvala}, {Kastner}, {Kov{\'a}cs}, {Lin}, {Loewenstein},
  {Looney}, {Lopez-Rodriguez}, {Maher}, {Michail}, {Miller}, {Moseley},
  {Novak}, {Pernic}, {Rennick}, {Rhody}, {Sandberg}, {Sandford}, {Santos},
  {Shafer}, {Sharp}, {Shirron}, {Siah}, {Silverberg}, {Sparr}, {Spotz},
  {Staguhn}, {Toorian}, {Towey}, {Tuttle}, {Vaillancourt}, {Voellmer},
  {Volpert}, {Wang}, \& {Wollack}}]{2018JAI.....740008H}
{Harper}, D.~A., {Runyan}, M.~C., {Dowell}, C.~D., {et~al.} 2018, Journal of
  Astronomical Instrumentation, 7, 1840008, \dodoi{10.1142/S2251171718400081}

\bibitem[{{Higuchi} {et~al.}(2015){Higuchi}, {Hasegawa}, {Saigo}, {Sanhueza},
  \& {Chibueze}}]{2015ApJ...815..106H}
{Higuchi}, A.~E., {Hasegawa}, T., {Saigo}, K., {Sanhueza}, P., \& {Chibueze},
  J.~O. 2015, \apj, 815, 106, \dodoi{10.1088/0004-637X/815/2/106}

\bibitem[{{Hildebrand}(1983)}]{1983QJRAS..24..267H}
{Hildebrand}, R.~H. 1983, \qjras, 24, 267

\bibitem[{{Hildebrand} {et~al.}(2009){Hildebrand}, {Kirby}, {Dotson}, {Houde},
  \& {Vaillancourt}}]{2009ApJ...696..567H}
{Hildebrand}, R.~H., {Kirby}, L., {Dotson}, J.~L., {Houde}, M., \&
  {Vaillancourt}, J.~E. 2009, \apj, 696, 567,
  \dodoi{10.1088/0004-637X/696/1/567}

\bibitem[{{Houde} {et~al.}(2016){Houde}, {Hull}, {Plambeck}, {Vaillancourt}, \&
  {Hildebrand}}]{2016ApJ...820...38H}
{Houde}, M., {Hull}, C. L.~H., {Plambeck}, R.~L., {Vaillancourt}, J.~E., \&
  {Hildebrand}, R.~H. 2016, \apj, 820, 38, \dodoi{10.3847/0004-637X/820/1/38}

\bibitem[{{Houde} {et~al.}(2009){Houde}, {Vaillancourt}, {Hildebrand},
  {Chitsazzadeh}, \& {Kirby}}]{2009ApJ...706.1504H}
{Houde}, M., {Vaillancourt}, J.~E., {Hildebrand}, R.~H., {Chitsazzadeh}, S., \&
  {Kirby}, L. 2009, \apj, 706, 1504, \dodoi{10.1088/0004-637X/706/2/1504}

\bibitem[{{Hull} \& {Zhang}(2019)}]{2019FrASS...6....3H}
{Hull}, C. L.~H., \& {Zhang}, Q. 2019, Frontiers in Astronomy and Space
  Sciences, 6, 3, \dodoi{10.3389/fspas.2019.00003}

\bibitem[{{Huttemeister} {et~al.}(1993){Huttemeister}, {Wilson}, {Henkel}, \&
  {Mauersberger}}]{1993A&A...276..445H}
{Huttemeister}, S., {Wilson}, T.~L., {Henkel}, C., \& {Mauersberger}, R. 1993,
  \aap, 276, 445

\bibitem[{{Hwang} {et~al.}(2022){Hwang}, {Kim}, {Pattle}, {Lee}, {Koch},
  {Johnstone}, {Tomisaka}, {Whitworth}, {Furuya}, {Kang}, {Lyo}, {Chung},
  {Arzoumanian}, {Park}, {Kwon}, {Kim}, {Tamura}, {Kwon}, {Soam}, {Han},
  {Hoang}, {Kim}, {Onaka}, {Eswaraiah}, {Ward-Thompson}, {Liu}, {Tang}, {Chen},
  {Matsumura}, {Hoang}, {Chen}, {Le Gouellec}, {Kirchschlager}, {Poidevin},
  {Bastien}, {Qiu}, {Hasegawa}, {Lai}, {Byun}, {Cho}, {Choi}, {Choi}, {Choi},
  {Jeong}, {Kang}, {Kim}, {Kim}, {Lee}, {Lee}, {Lee}, {Lee}, {Kim}, {Yoo},
  {Yun}, {Chen}, {Di Francesco}, {Fiege}, {Fissel}, {Franzmann}, {Houde},
  {Lacaille}, {Matthews}, {Sadavoy}, {Moriarty-Schieven}, {Tahani}, {Ching},
  {Dai}, {Duan}, {Gu}, {Law}, {Li}, {Li}, {Li}, {Li}, {Liu}, {Lu}, {Qian},
  {Wang}, {Wu}, {Xie}, {Yuan}, {Zhang}, {Zhang}, {Zhang}, {Zhou}, {Zhu},
  {Berry}, {Friberg}, {Graves}, {Liu}, {Mairs}, {Parsons}, {Rawlings}, {Doi},
  {Hayashi}, {Hull}, {Inoue}, {Inutsuka}, {Iwasaki}, {Kataoka}, {Kawabata},
  {Kim}, {Kobayashi}, {Nagata}, {Nakamura}, {Nakanishi}, {Pyo}, {Saito},
  {Seta}, {Shimajiri}, {Shinnaga}, {Tsukamoto}, {Zenko}, {Chen}, {Duan},
  {Fanciullo}, {Kemper}, {Lee}, {Lin}, {Liu}, {Ohashi}, {Rao}, {Tang}, {Wang},
  {Yang}, {Yen}, {Bourke}, {Chrysostomou}, {Debattista}, {Eden}, {Eyres},
  {Falle}, {Fuller}, {Gledhill}, {Greaves}, {Griffin}, {Hatchell}, {Karoly},
  {Kirk}, {K{\"o}nyves}, {Longmore}, {van Loo}, {de Looze}, {Peretto},
  {Priestley}, {Rawlings}, {Retter}, {Richer}, {Rigby}, {Savini}, {Scaife},
  {Viti}, {Diep}, {Ngoc}, {Tram}, {Andr{\'e}}, {Coud{\'e}}, {Dowell},
  {Friesen}, \& {Robitaille}}]{2022ApJ...941...51H}
{Hwang}, J., {Kim}, J., {Pattle}, K., {et~al.} 2022, \apj, 941, 51,
  \dodoi{10.3847/1538-4357/ac99e0}

\bibitem[{{Jones} {et~al.}(2008){Jones}, {Burton}, {Cunningham}, {Menten},
  {Schilke}, {Belloche}, {Leurini}, {Ott}, \& {Walsh}}]{2008MNRAS.386..117J}
{Jones}, P.~A., {Burton}, M.~G., {Cunningham}, M.~R., {et~al.} 2008, \mnras,
  386, 117, \dodoi{10.1111/j.1365-2966.2008.13009.x}

\bibitem[{{Kauffmann} {et~al.}(2017){Kauffmann}, {Pillai}, {Zhang}, {Menten},
  {Goldsmith}, {Lu}, \& {Guzm{\'a}n}}]{2017A&A...603A..89K}
{Kauffmann}, J., {Pillai}, T., {Zhang}, Q., {et~al.} 2017, \aap, 603, A89,
  \dodoi{10.1051/0004-6361/201628088}

\bibitem[{{Kirk} {et~al.}(2013){Kirk}, {Ward-Thompson}, {Palmeirim},
  {Andr{\'e}}, {Griffin}, {Hargrave}, {K{\"o}nyves}, {Bernard}, {Nutter},
  {Sibthorpe}, {Di Francesco}, {Abergel}, {Arzoumanian}, {Benedettini},
  {Bontemps}, {Elia}, {Hennemann}, {Hill}, {Men'shchikov}, {Motte},
  {Nguyen-Luong}, {Peretto}, {Pezzuto}, {Rygl}, {Sadavoy}, {Schisano},
  {Schneider}, {Testi}, \& {White}}]{2013MNRAS.432.1424K}
{Kirk}, J.~M., {Ward-Thompson}, D., {Palmeirim}, P., {et~al.} 2013, \mnras,
  432, 1424, \dodoi{10.1093/mnras/stt561}

\bibitem[{{Koch} {et~al.}(2012){Koch}, {Tang}, \& {Ho}}]{2012ApJ...747...79K}
{Koch}, P.~M., {Tang}, Y.-W., \& {Ho}, P. T.~P. 2012, \apj, 747, 79,
  \dodoi{10.1088/0004-637X/747/1/79}

\bibitem[{{Krumholz} \& {Federrath}(2019)}]{2019FrASS...6....7K}
{Krumholz}, M.~R., \& {Federrath}, C. 2019, Frontiers in Astronomy and Space
  Sciences, 6, 7, \dodoi{10.3389/fspas.2019.00007}

\bibitem[{{Law} {et~al.}(2008){Law}, {Yusef-Zadeh}, \&
  {Cotton}}]{2008ApJS..177..515L}
{Law}, C.~J., {Yusef-Zadeh}, F., \& {Cotton}, W.~D. 2008, \apjs, 177, 515,
  \dodoi{10.1086/588218}

\bibitem[{{Lazarian}(2007)}]{2007JQSRT.106..225L}
{Lazarian}, A. 2007, \jqsrt, 106, 225, \dodoi{10.1016/j.jqsrt.2007.01.038}

\bibitem[{{Lazarian} \& {Hoang}(2007)}]{2007MNRAS.378..910L}
{Lazarian}, A., \& {Hoang}, T. 2007, \mnras, 378, 910,
  \dodoi{10.1111/j.1365-2966.2007.11817.x}

\bibitem[{{Li} {et~al.}(2015){Li}, {Yuen}, {Otto}, {Leung}, {Sridharan},
  {Zhang}, {Liu}, {Tang}, \& {Qiu}}]{2015Natur.520..518L}
{Li}, H.-B., {Yuen}, K.~H., {Otto}, F., {et~al.} 2015, \nat, 520, 518,
  \dodoi{10.1038/nature14291}

\bibitem[{{Lis} \& {Goldsmith}(1991)}]{1991ApJ...369..157L}
{Lis}, D.~C., \& {Goldsmith}, P.~F. 1991, \apj, 369, 157,
  \dodoi{10.1086/169746}

\bibitem[{{Liu} {et~al.}(2021){Liu}, {Zhang}, {Commer{\c{c}}on}, {Valdivia},
  {Maury}, \& {Qiu}}]{2021ApJ...919...79L}
{Liu}, J., {Zhang}, Q., {Commer{\c{c}}on}, B., {et~al.} 2021, \apj, 919, 79,
  \dodoi{10.3847/1538-4357/ac0cec}

\bibitem[{{Liu} {et~al.}(2022){Liu}, {Zhang}, \& {Qiu}}]{2022FrASS...9.3556L}
{Liu}, J., {Zhang}, Q., \& {Qiu}, K. 2022, Frontiers in Astronomy and Space
  Sciences, 9, 943556, \dodoi{10.3389/fspas.2022.943556}

\bibitem[{{Liu} {et~al.}(2020){Liu}, {Zhang}, {Qiu}, {Liu}, {Pillai}, {Girart},
  {Li}, \& {Wang}}]{2020ApJ...895..142L}
{Liu}, J., {Zhang}, Q., {Qiu}, K., {et~al.} 2020, \apj, 895, 142,
  \dodoi{10.3847/1538-4357/ab9087}

\bibitem[{{Liu} {et~al.}(2023){Liu}, {Zhang}, {Koch}, {Liu}, {Li}, {Li},
  {Girart}, {Chen}, {Ching}, {Ho}, {Lai}, {Qiu}, {Rao}, \&
  {Tang}}]{2023ApJ...945..160L}
{Liu}, J., {Zhang}, Q., {Koch}, P.~M., {et~al.} 2023, \apj, 945, 160,
  \dodoi{10.3847/1538-4357/acb540}

\bibitem[{{Longmore} {et~al.}(2013){Longmore}, {Bally}, {Testi}, {Purcell},
  {Walsh}, {Bressert}, {Pestalozzi}, {Molinari}, {Ott}, {Cortese}, {Battersby},
  {Murray}, {Lee}, {Kruijssen}, {Schisano}, \& {Elia}}]{2013MNRAS.429..987L}
{Longmore}, S.~N., {Bally}, J., {Testi}, L., {et~al.} 2013, \mnras, 429, 987,
  \dodoi{10.1093/mnras/sts376}

\bibitem[{{Lopez-Rodriguez}(2023)}]{2023ApJ...953..113L}
{Lopez-Rodriguez}, E. 2023, \apj, 953, 113, \dodoi{10.3847/1538-4357/ace110}

\bibitem[{{Lopez-Rodriguez} {et~al.}(2021){Lopez-Rodriguez}, {Guerra},
  {Asgari-Targhi}, \& {Schmelz}}]{2021ApJ...914...24L}
{Lopez-Rodriguez}, E., {Guerra}, J.~A., {Asgari-Targhi}, M., \& {Schmelz},
  J.~T. 2021, \apj, 914, 24, \dodoi{10.3847/1538-4357/abf934}

\bibitem[{{Louvet} {et~al.}(2014){Louvet}, {Motte}, {Hennebelle}, {Maury},
  {Bonnell}, {Bontemps}, {Gusdorf}, {Hill}, {Gueth}, {Peretto},
  {Duarte-Cabral}, {Stephan}, {Schilke}, {Csengeri}, {Nguyen Luong}, \&
  {Lis}}]{2014A&A...570A..15L}
{Louvet}, F., {Motte}, F., {Hennebelle}, P., {et~al.} 2014, \aap, 570, A15,
  \dodoi{10.1051/0004-6361/201423603}

\bibitem[{{Lu} {et~al.}(2019){Lu}, {Mills}, {Ginsburg}, {Walker}, {Barnes},
  {Butterfield}, {Henshaw}, {Battersby}, {Kruijssen}, {Longmore}, {Zhang},
  {Bally}, {Kauffmann}, {Ott}, {Rickert}, \& {Wang}}]{2019ApJS..244...35L}
{Lu}, X., {Mills}, E. A.~C., {Ginsburg}, A., {et~al.} 2019, \apjs, 244, 35,
  \dodoi{10.3847/1538-4365/ab4258}

\bibitem[{{Marrone}(2006)}]{2006PhDT........32M}
{Marrone}, D.~P. 2006, PhD thesis, Harvard University, Massachusetts

\bibitem[{{McGrath} {et~al.}(2004){McGrath}, {Goss}, \& {De
  Pree}}]{2004ApJS..155..577M}
{McGrath}, E.~J., {Goss}, W.~M., \& {De Pree}, C.~G. 2004, \apjs, 155, 577,
  \dodoi{10.1086/424486}

\bibitem[{{Mehringer} {et~al.}(1995){Mehringer}, {Palmer}, {Goss}, \&
  {Yusef-Zadeh}}]{1995ADIL...DM...02M}
{Mehringer}, D.~M., {Palmer}, P., {Goss}, W.~M., \& {Yusef-Zadeh}, F. 1995,
  Astronomy Data Image Library

\bibitem[{{Menon} {et~al.}(2020){Menon}, {Federrath}, \&
  {Kuiper}}]{2020MNRAS.493.4643M}
{Menon}, S.~H., {Federrath}, C., \& {Kuiper}, R. 2020, \mnras, 493, 4643,
  \dodoi{10.1093/mnras/staa580}

\bibitem[{{Motte} {et~al.}(2012){Motte}, {Bontemps}, {Hennemann}, {Nguyen
  Luong}, {Schneider}, {Didelon}, \& {Zavagno}}]{2012sf2a.conf...45M}
{Motte}, F., {Bontemps}, S., {Hennemann}, M., {et~al.} 2012, in SF2A-2012:
  Proceedings of the Annual meeting of the French Society of Astronomy and
  Astrophysics, ed. S.~{Boissier}, P.~{de Laverny}, N.~{Nardetto}, R.~{Samadi},
  D.~{Valls-Gabaud}, \& H.~{Wozniak}, 45--50

\bibitem[{{Motte} {et~al.}(2003){Motte}, {Schilke}, \&
  {Lis}}]{2003ApJ...582..277M}
{Motte}, F., {Schilke}, P., \& {Lis}, D.~C. 2003, \apj, 582, 277,
  \dodoi{10.1086/344538}

\bibitem[{{Myers} {et~al.}(2022){Myers}, {Hatchfield}, \&
  {Battersby}}]{2022ApJ...929...34M}
{Myers}, P.~C., {Hatchfield}, H.~P., \& {Battersby}, C. 2022, \apj, 929, 34,
  \dodoi{10.3847/1538-4357/ac5906}

\bibitem[{{Nakano} \& {Nakamura}(1978)}]{1978PASJ...30..671N}
{Nakano}, T., \& {Nakamura}, T. 1978, \pasj, 30, 671

\bibitem[{{Neill} {et~al.}(2014){Neill}, {Bergin}, {Lis}, {Schilke},
  {Crockett}, {Favre}, {Emprechtinger}, {Comito}, {Qin}, {Anderson},
  {Burkhardt}, {Chen}, {Harris}, {Lord}, {McGuire}, {McNeill}, {Monje},
  {Phillips}, {Steber}, {Vasyunina}, \& {Yu}}]{2014ApJ...789....8N}
{Neill}, J.~L., {Bergin}, E.~A., {Lis}, D.~C., {et~al.} 2014, \apj, 789, 8,
  \dodoi{10.1088/0004-637X/789/1/8}

\bibitem[{{Novak} {et~al.}(1997){Novak}, {Dotson}, {Dowell}, {Goldsmith},
  {Hildebrand}, {Platt}, \& {Schleuning}}]{1997ApJ...487..320N}
{Novak}, G., {Dotson}, J.~L., {Dowell}, C.~D., {et~al.} 1997, \apj, 487, 320,
  \dodoi{10.1086/304576}

\bibitem[{{Ostriker} {et~al.}(2001){Ostriker}, {Stone}, \&
  {Gammie}}]{2001ApJ...546..980O}
{Ostriker}, E.~C., {Stone}, J.~M., \& {Gammie}, C.~F. 2001, \apj, 546, 980,
  \dodoi{10.1086/318290}

\bibitem[{{Padoan} {et~al.}(2001){Padoan}, {Goodman}, {Draine}, {Juvela},
  {Nordlund}, \& {R{\"o}gnvaldsson}}]{2001ApJ...559.1005P}
{Padoan}, P., {Goodman}, A., {Draine}, B.~T., {et~al.} 2001, \apj, 559, 1005,
  \dodoi{10.1086/322504}

\bibitem[{{Par{\'e}} {et~al.}(2024){Par{\'e}}, {Butterfield}, {Chuss},
  {Guerra}, {Iuliano}, {Karpovich}, {Morris}, \&
  {Wollack}}]{2024arXiv240105317P}
{Par{\'e}}, D., {Butterfield}, N.~O., {Chuss}, D.~T., {et~al.} 2024, arXiv
  e-prints, arXiv:2401.05317, \dodoi{10.48550/arXiv.2401.05317}

\bibitem[{{Pattle} {et~al.}(2023){Pattle}, {Fissel}, {Tahani}, {Liu}, \&
  {Ntormousi}}]{2023ASPC..534..193P}
{Pattle}, K., {Fissel}, L., {Tahani}, M., {Liu}, T., \& {Ntormousi}, E. 2023,
  in Astronomical Society of the Pacific Conference Series, Vol. 534,
  Protostars and Planets VII, ed. S.~{Inutsuka}, Y.~{Aikawa}, T.~{Muto},
  K.~{Tomida}, \& M.~{Tamura}, 193, \dodoi{10.48550/arXiv.2203.11179}

\bibitem[{{Pattle} {et~al.}(2015){Pattle}, {Ward-Thompson}, {Kirk}, {White},
  {Drabek-Maunder}, {Buckle}, {Beaulieu}, {Berry}, {Broekhoven-Fiene},
  {Currie}, {Fich}, {Hatchell}, {Kirk}, {Jenness}, {Johnstone}, {Mottram},
  {Nutter}, {Pineda}, {Quinn}, {Salji}, {Tisi}, {Walker-Smith}, {di Francesco},
  {Hogerheijde}, {Andr{\'e}}, {Bastien}, {Bresnahan}, {Butner}, {Chen},
  {Chrysostomou}, {Coude}, {Davis}, {Duarte-Cabral}, {Fiege}, {Friberg},
  {Friesen}, {Fuller}, {Graves}, {Greaves}, {Gregson}, {Griffin}, {Holland},
  {Joncas}, {Knee}, {K{\"o}nyves}, {Mairs}, {Marsh}, {Matthews},
  {Moriarty-Schieven}, {Rawlings}, {Richer}, {Robertson}, {Rosolowsky},
  {Rumble}, {Sadavoy}, {Spinoglio}, {Thomas}, {Tothill}, {Viti}, {Wouterloot},
  {Yates}, \& {Zhu}}]{2015MNRAS.450.1094P}
{Pattle}, K., {Ward-Thompson}, D., {Kirk}, J.~M., {et~al.} 2015, \mnras, 450,
  1094, \dodoi{10.1093/mnras/stv376}

\bibitem[{{Persi} \& {Tapia}(2008)}]{2008hsf2.book..456P}
{Persi}, P., \& {Tapia}, M. 2008, in Handbook of Star Forming Regions, Volume
  II, ed. B.~{Reipurth}, Vol.~5, 456

\bibitem[{{Pierce-Price} {et~al.}(2000){Pierce-Price}, {Richer}, {Greaves},
  {Holland}, {Jenness}, {Lasenby}, {White}, {Matthews}, {Ward-Thompson},
  {Dent}, {Zylka}, {Mezger}, {Hasegawa}, {Oka}, {Omont}, \&
  {Gilmore}}]{2000ApJ...545L.121P}
{Pierce-Price}, D., {Richer}, J.~S., {Greaves}, J.~S., {et~al.} 2000, \apjl,
  545, L121, \dodoi{10.1086/317884}

\bibitem[{{Qin} {et~al.}(2011){Qin}, {Schilke}, {Rolffs}, {Comito}, {Lis}, \&
  {Zhang}}]{2011A&A...530L...9Q}
{Qin}, S.~L., {Schilke}, P., {Rolffs}, R., {et~al.} 2011, \aap, 530, L9,
  \dodoi{10.1051/0004-6361/201116928}

\bibitem[{{Qin} {et~al.}(2008){Qin}, {Zhao}, {Moran}, {Marrone}, {Patel},
  {Wang}, {Liu}, \& {Kuan}}]{2008ApJ...677..353Q}
{Qin}, S.-L., {Zhao}, J.-H., {Moran}, J.~M., {et~al.} 2008, \apj, 677, 353,
  \dodoi{10.1086/529067}

\bibitem[{{Qiu} {et~al.}(2014){Qiu}, {Zhang}, {Menten}, {Liu}, {Tang}, \&
  {Girart}}]{2014ApJ...794L..18Q}
{Qiu}, K., {Zhang}, Q., {Menten}, K.~M., {et~al.} 2014, \apjl, 794, L18,
  \dodoi{10.1088/2041-8205/794/1/L18}

\bibitem[{{Reid} {et~al.}(2014){Reid}, {Menten}, {Brunthaler}, {Zheng}, {Dame},
  {Xu}, {Wu}, {Zhang}, {Sanna}, {Sato}, {Hachisuka}, {Choi}, {Immer},
  {Moscadelli}, {Rygl}, \& {Bartkiewicz}}]{2014ApJ...783..130R}
{Reid}, M.~J., {Menten}, K.~M., {Brunthaler}, A., {et~al.} 2014, \apj, 783,
  130, \dodoi{10.1088/0004-637X/783/2/130}

\bibitem[{{Russeil} {et~al.}(2017){Russeil}, {Adami}, {Bouret}, {Herv{\'e}},
  {Parker}, {Zavagno}, \& {Motte}}]{2017A&A...607A..86R}
{Russeil}, D., {Adami}, C., {Bouret}, J.~C., {et~al.} 2017, \aap, 607, A86,
  \dodoi{10.1051/0004-6361/201629870}

\bibitem[{{Sanhueza} {et~al.}(2021){Sanhueza}, {Girart}, {Padovani}, {Galli},
  {Hull}, {Zhang}, {Cortes}, {Stephens}, {Fern{\'a}ndez-L{\'o}pez}, {Jackson},
  {Frau}, {Kock}, {Wu}, {Zapata}, {Olguin}, {Lu}, {Silva}, {Tang}, {Sakai},
  {Guzm{\'a}n}, {Tatematsu}, {Nakamura}, \& {Chen}}]{2021ApJ...915L..10S}
{Sanhueza}, P., {Girart}, J.~M., {Padovani}, M., {et~al.} 2021, \apjl, 915,
  L10, \dodoi{10.3847/2041-8213/ac081c}

\bibitem[{{Schmiedeke} {et~al.}(2016){Schmiedeke}, {Schilke}, {M{\"o}ller},
  {S{\'a}nchez-Monge}, {Bergin}, {Comito}, {Csengeri}, {Lis}, {Molinari},
  {Qin}, \& {Rolffs}}]{2016A&A...588A.143S}
{Schmiedeke}, A., {Schilke}, P., {M{\"o}ller}, T., {et~al.} 2016, \aap, 588,
  A143, \dodoi{10.1051/0004-6361/201527311}

\bibitem[{{Schw{\"o}rer} {et~al.}(2019){Schw{\"o}rer}, {S{\'a}nchez-Monge},
  {Schilke}, {M{\"o}ller}, {Ginsburg}, {Meng}, {Schmiedeke}, {M{\"u}ller},
  {Lis}, \& {Qin}}]{2019A&A...628A...6S}
{Schw{\"o}rer}, A., {S{\'a}nchez-Monge}, {\'A}., {Schilke}, P., {et~al.} 2019,
  \aap, 628, A6, \dodoi{10.1051/0004-6361/201935200}

\bibitem[{{Sridharan} {et~al.}(2014){Sridharan}, {Rao}, {Qiu}, {Cortes}, {Li},
  {Pillai}, {Patel}, \& {Zhang}}]{2014ApJ...783L..31S}
{Sridharan}, T.~K., {Rao}, R., {Qiu}, K., {et~al.} 2014, \apjl, 783, L31,
  \dodoi{10.1088/2041-8205/783/2/L31}

\bibitem[{{Tan} {et~al.}(2014){Tan}, {Beltr{\'a}n}, {Caselli}, {Fontani},
  {Fuente}, {Krumholz}, {McKee}, \& {Stolte}}]{2014prpl.conf..149T}
{Tan}, J.~C., {Beltr{\'a}n}, M.~T., {Caselli}, P., {et~al.} 2014, in Protostars
  and Planets VI, ed. H.~{Beuther}, R.~S. {Klessen}, C.~P. {Dullemond}, \&
  T.~{Henning}, 149--172, \dodoi{10.2458/azu_uapress_9780816531240-ch007}

\bibitem[{{Temi} {et~al.}(2018){Temi}, {Hoffman}, {Ennico}, \&
  {Le}}]{2018JAI.....740011T}
{Temi}, P., {Hoffman}, D., {Ennico}, K., \& {Le}, J. 2018, Journal of
  Astronomical Instrumentation, 7, 1840011, \dodoi{10.1142/S2251171718400111}

\bibitem[{{Vaillancourt}(2006)}]{2006PASP..118.1340V}
{Vaillancourt}, J.~E. 2006, \pasp, 118, 1340, \dodoi{10.1086/507472}

\bibitem[{{Walsh} {et~al.}(2014){Walsh}, {Purcell}, {Longmore}, {Breen},
  {Green}, {Harvey-Smith}, {Jordan}, \& {Macpherson}}]{2014MNRAS.442.2240W}
{Walsh}, A.~J., {Purcell}, C.~R., {Longmore}, S.~N., {et~al.} 2014, \mnras,
  442, 2240, \dodoi{10.1093/mnras/stu989}

\bibitem[{{Willis} {et~al.}(2013){Willis}, {Marengo}, {Allen}, {Fazio},
  {Smith}, \& {Carey}}]{2013ApJ...778...96W}
{Willis}, S., {Marengo}, M., {Allen}, L., {et~al.} 2013, \apj, 778, 96,
  \dodoi{10.1088/0004-637X/778/2/96}

\bibitem[{{Zeng} {et~al.}(2023){Zeng}, {Zhang}, {Alves}, {Ching}, {Girart}, \&
  {Liu}}]{2023ApJ...954...99Z}
{Zeng}, L., {Zhang}, Q., {Alves}, F.~O., {et~al.} 2023, \apj, 954, 99,
  \dodoi{10.3847/1538-4357/ace690}

\bibitem[{{Zhang} {et~al.}(2014){Zhang}, {Qiu}, {Girart}, {Liu}, {Tang},
  {Koch}, {Li}, {Keto}, {Ho}, {Rao}, {Lai}, {Ching}, {Frau}, {Chen}, {Li},
  {Padovani}, {Bontemps}, {Csengeri}, \& {Ju{\'a}rez}}]{2014ApJ...792..116Z}
{Zhang}, Q., {Qiu}, K., {Girart}, J.~M., {et~al.} 2014, \apj, 792, 116,
  \dodoi{10.1088/0004-637X/792/2/116}

\end{thebibliography}
\bibliographystyle{aasjournal}

%% This command is needed to show the entire author+affiliation list when
%% the collaboration and author truncation commands are used.  It has to
%% go at the end of the manuscript.
%\allauthors

%% Include this line if you are using the \added, \replaced, \deleted
%% commands to see a summary list of all changes at the end of the article.
%\listofchanges

\end{document}